# Multiple-scale structures: from Faraday waves to soft-matter quasicrystals

Samuel Savitz,[a]* Mehrtash Babadi[b,a] and Ron Lifshitz[c,a]*

[a]Condensed Matter Physics, California Institute of Technology, Pasadena, CA 91125, USA, [b]Broad Institute of MIT and Harvard, Cambridge, MA 02142, USA, and [c]Raymond and Beverly Sackler School of Physics and Astronomy, Tel Aviv University, Tel Aviv 69978, Israel. *Correspondence e-mail: sam@savitz.org, ronlif@tau.ac.il

For many years, quasicrystals were observed only as solid-state metallic alloys, yet current research is now actively exploring their formation in a variety of soft materials, including systems of macromolecules, nanoparticles and colloids. Much effort is being invested in understanding the thermodynamic properties of these soft-matter quasicrystals in order to predict and possibly control the structures that form, and hopefully to shed light on the broader yet unresolved general questions of quasicrystal formation and stability. Moreover, the ability to control the self-assembly of soft quasicrystals may contribute to the development of novel photonics or other applications based on self-assembled metamaterials. Here a path is followed, leading to quantitative stability predictions, that starts with a model developed two decades ago to treat the formation of multiple-scale quasiperiodic Faraday waves (standing wave patterns in vibrating fluid surfaces) and which was later mapped onto systems of soft particles, interacting via multiple-scale pair potentials. The article reviews, and substantially expands, the quantitative predictions of these models, while correcting a few discrepancies in earlier calculations, and presents new analytical methods for treating the models. In so doing, a number of new stable quasicrystalline structures are found with octagonal, octadecagonal and higher-order symmetries, some of which may, it is hoped, be observed in future experiments.

## 1. Introduction and outline

The scope of research on quasicrystals[1] has greatly expanded in the last decade, owing mainly to the advent of a host of new experimental systems exhibiting aperiodic structures with long-range order. The ever-growing variety of stable *solid-state* quasicrystals (Tsai, 2003, 2008; Janssen *et al.*, 2007; Steurer & Deloudi, 2009), where quasiperiodic long-range order occurs on the atomic scale, has been joined in recent years by a host of exciting new *soft-matter* systems that exhibit this quasiperiodicity on a larger mesoscopic scale – typically from a few nanometres to a few micrometres (Zeng *et al.*, 2004; Ungar & Zeng, 2005; Takano *et al.*, 2005; Hayashida *et al.*, 2007; Percec *et al.*, 2009; Talapin *et al.*, 2009; Ungar *et al.*, 2011; Dotera, 2011, 2012; Fischer *et al.*, 2011; Xiao *et al.*, 2012; Zhang & Bates, 2012; Bodnarchuk *et al.*, 2013; Chanpuriya *et al.*, 2016). In addition to having promising applications, particularly as metamaterials in the optical domain, these substances give us the opportunity to study quasicrystals in ways that were impossible before. The obvious reason for this is indeed the fact that their building blocks – rather than being individual atoms – are composed of large synthesized particles such as macromolecules, nanoparticles and colloids. At these dimensions it may be possible to track the dynamics of individual particles, manipulate their positions or possibly design the interaction between them. If so, an obvious question to ask is how to design this interaction to obtain a particular desired quasicrystal. To answer this question, one clearly requires an understanding of the spontaneous formation and subsequent stability of such materials.

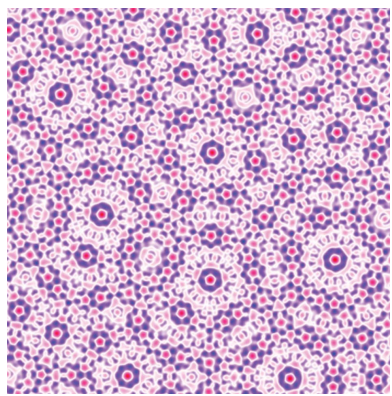

---

[1] For definitions, basic terminology and some background on the symmetry-breaking transition from a liquid phase to a quasicrystal see, for example, Lifshitz (2003, 2007, 2011).

Phenomenological Landau theories based on *ad hoc* free energies have been widely applied to study the thermodynamics of phase transitions (Alexander & McTague, 1978) and to explain the stability of different phases, including quasicrystals (Mermin & Troian, 1985; Bak, 1985; Kalugin *et al.*, 1985; Jaric, 1985; Gronlund & Mermin, 1988; Narasimhan & Ho, 1988). In such models, one identifies an order parameter field – which is often a simple scalar function $\rho(\mathbf{r})$ that describes the relative deviation $[c(\mathbf{r}) - \overline{c}]/\overline{c}$ of a coarse-grained density $c(\mathbf{r})$ of a material from its average value $\overline{c}$ – and uses generic symmetry arguments to formulate a free-energy functional $\mathcal{F}[\rho(\mathbf{r})]$, expressed in powers of the order parameter and its gradients. One assumes that the equilibrium phase is the one that minimizes $\mathcal{F}$, and then all that remains is to find out which structures could minimize such a free energy, or conversely, how to tweak $\mathcal{F}$ to obtain the desired structures. For a textbook review see, for example, chapter 4 of the book by Chaikin & Lubensky (1995). These models are particularly attractive for soft-matter systems (de Gennes & Prost, 1993; Gompper & Schick, 1994) owing to their mesoscopic building blocks, which render the long-wavelength gradient expansion and the truncation at low order a more valid approximation than in the atomic case, especially when the transition is only weakly first-order.

Important insight into the stability question emerged when Edwards & Fauve (1993) discovered that parametrically driven liquid surfaces, exhibiting standing-wave patterns known as Faraday waves, can become quasiperiodic when driven by a superposition of two temporal frequencies [see also Kudrolli *et al.* (1998), Gollub & Langer (1999) and Arbell & Fineberg (2002)]. The realization that these temporal frequencies impose two spatial length scales on the stable structures that form prompted Lifshitz & Petrich (1997), henceforth LP, to generalize the Swift–Hohenberg equation (Swift & Hohenberg, 1977) and introduce a rather simple Landau free-energy expansion (or a Lyapunov functional) of a scalar field $\rho(\mathbf{r})$ in two dimensions of the form

$$\mathcal{F}_{\mathrm{LP}} = \frac{c}{2} \int \left[ \left(\nabla^2 + 1\right)\left(\nabla^2 + q^2\right)\rho(\mathbf{r}) \right]^2 \mathrm{d}\mathbf{r} + \int \left\{ -\frac{\epsilon}{2} \rho(\mathbf{r})^2 - \frac{\alpha}{3} \rho(\mathbf{r})^3 + \frac{1}{4} \rho(\mathbf{r})^4 \right\} \mathrm{d}\mathbf{r}. \quad (1)$$

In Section 2 we carefully review the basic features of this free energy. Here, we only wish to point out its two main features:

(i) The first integral in $\mathcal{F}_{\mathrm{LP}}$, containing the non-local gradient expansion, is responsible for selecting two length scales, whose ratio is given by the parameter $q$. This is because density modes with wavenumbers differing from unity or $q$ increase its value.

(ii) The second integral, containing the local expansion in powers of the order parameter field, contains an odd, cubic, power that breaks the $\mathbb{Z}_2$ symmetry of $\rho \to -\rho$. It is this term that is responsible for stabilizing structures with two length scales, depending on the value of $q$, as it has the ability to lower the free energy if there exist triplets of density modes with wavevectors that add up to zero. These are known in the Faraday wave literature as triad resonances, and amount to effective three-body interaction in the coarse-grained density context.

In particular, by setting the value of the wavenumber ratio $q$ to

$$k_n = 2\cos\frac{\pi}{n}, \quad (2)$$

one can form triplets or 'triangles' containing two unit wavevectors separated by $2\pi/n$ and a third wavevector of length $k_n$. This may sufficiently lower the free energy of structures with $N$-fold rotational symmetry (where $N$ is equal to $n$ or $2n$ for even or odd $n$, respectively), making them the ability to give the absolute minimum of $\mathcal{F}_{\mathrm{LP}}$. In Section 3 we repeat and extend the calculations of LP of the free energies of candidate structures, setting $q = k_n$ for different values of $n$ and assuming LP's limit of $c \to \infty$, which leads to exact length-scale selectivity. In doing so we provide a more complete and definitive calculation, while correcting a few discrepancies in their results that have caused some confusion over the years.

Owing to its simplicity and clarity in explaining the stability of the decagonal (10-fold) and dodecagonal (12-fold) quasicrystals that it exhibits, as well as the ease with which one can numerically simulate the dynamical equation that it generates *via* simple relaxation $\partial_t \rho = -\delta\mathcal{F}_{\mathrm{LP}}/\delta\rho$, the LP model has been studied in depth since its original publication and extended in a number of different ways (Wu *et al.*, 2010; Mkhonta *et al.*, 2013; Achim *et al.*, 2014; Jiang & Zhang, 2014; Jiang *et al.*, 2015, 2016, 2017; Subramanian *et al.*, 2016).

Here we further extend the LP model as follows:

(i) Jiang & Zhang (2014) and Jiang *et al.* (2015) improved on the free-energy calculations of LP by relaxing the exact length selection imposed by the $c \to \infty$ limit, using a high-dimensional numerical evaluation scheme which they call the 'projection method'. In Section 4 we introduce an approximation scheme for calculating the LP free energy (1) with finite $c$ that allows competing structures to contain two length scales that are roughly, rather than exactly, equal to unity or $q$, thus improving their competitiveness and reducing the regions in parameter space where the quasicrystalline structures are stable. This qualitatively captures the importance of length-scale selectivity, but is quantitatively accurate only in the dodecagonal case. Nevertheless, it is much simpler to evaluate and provides further analytical insight about the model.

(ii) The only quasicrystals which can be stabilized by the original LP model, which allows for two length scales in the structures, are the decagonal and dodecagonal phases. In Section 5, we show that increasing the number of allowed length scales from two to four allows for the stabilization of octagonal (8-fold) and octadecagonal (18-fold) quasicrystals.

One can improve on the Landau expansions by using a density functional mean-field theory (Ramakrishnan & Yussouff, 1979), which is valid for all orders, by rigorously coarse-graining a system of interacting discrete particles. For a textbook introduction see, for example, chapter 5 of the book by Fredrickson (2006). Such theories were also considered in the early studies of quasicrystals (Sachdev & Nelson, 1985). A particularly simple coarse-grained free-energy functional of the form

$$\mathcal{F}_{CG} = \frac{1}{2} \iint c(\mathbf{r}) \mathcal{U}(\mathbf{r} - \mathbf{r}') c(\mathbf{r}') \, d\mathbf{r} \, d\mathbf{r}'$$
$$+ \int \{ k_B T c(\mathbf{r}) [\ln c(\mathbf{r}) - 1] - \mu c(\mathbf{r}) \} \, d\mathbf{r}, \qquad (3)$$

containing the familiar mean-field terms of pair interaction and ideal entropy, was used by Barkan, Diamant & Lifshitz (2011), henceforth BDL, as an extension of the LP model, to study the stability of soft-matter quasicrystals, initially in two dimensions. Again, one assumes that the equilibrium density field is the one that minimizes $\mathcal{F}_{CG}$ for the given thermodynamic parameters – such as temperature $T$ and chemical potential $\mu$ – and is then left with the question of how to design the pair potential $\mathcal{U}(\mathbf{r})$ to obtain the desired structures, giving us the ability to address our starting question.

To do so, BDL followed an earlier conjecture of Lifshitz & Diamant (2007), who attributed the stability of certain soft quasicrystals to the same mechanism that stabilizes the Faraday wave structures, namely the existence of two length scales in the pair potential, combined with effective many-body interactions. That stable quasicrystals may require the existence of two length scales in their effective interaction potentials $\mathcal{U}(\mathbf{r})$ is not a new idea (Olami, 1990; Smith, 1991). Many two-length-scale potentials have been investigated numerically over the years and found to exhibit stable quasi-periodic phases (Dzugutov, 1993; Jagla, 1998; Skibinsky et al., 1999; Quandt & Teter, 1999; Roth & Denton, 2000; Engel & Trebin, 2007; Keys & Glotzer, 2007; Archer et al., 2013; Dotera et al., 2014; Engel et al., 2015; Pattabhiraman & Dijkstra, 2017a; Damasceno et al., 2017). The novelty and emphasis of BDL was in their quantitative understanding of the stabilization mechanism – comparing the nonlinear pairwise interaction and the local entropy terms of $\mathcal{F}_{CG}$ to the nonlocal gradient expansion and local power expansion terms of $\mathcal{F}_{LP}$, respectively. This allowed them to pinpoint regions of stability in the parameter spaces of different potentials instead of performing exhaustive searches. Indeed, Barkan et al. (2014) confirmed these predictions by employing molecular dynamics simulations with particles that interact through pair potentials that were designed according to the principles of BDL. By properly setting the two length scales in these potentials, they were able to generate periodic crystals with square and hexagonal symmetry, quasicrystals with decagonal and do-decagonal symmetry, and a lamellar (or striped) phase.

The inclusion of a second length scale in $\mathcal{U}(\mathbf{r})$, imitating the gradient term of $\mathcal{F}_{LP}$, provides greater control over the self-assembly of desired structures than can be achieved with just a single scale, and turns out to be the key to obtaining stable quasicrystals and other novel structures. Yet, calculating the exact value of the coarse-grained free energy $\mathcal{F}_{CG}$ turned out to be a challenge. Instead, BDL expanded the logarithmic entropy term in a power series to fourth order in $\rho(\mathbf{r}) = [c(\mathbf{r}) - \bar{c}]/\bar{c}$ and mapped the resulting approximate free energy onto the LP free energy, thus obtaining a rough estimate of the physical parameters that stabilize the different structures using the results known for the LP model.

Here we present new insight into the stability of soft-matter quasicrystals, by significantly improving upon the original BDL analysis as follows:

(i) In Section 6, we introduce the 'density distribution method' for evaluating the free energy of candidate structures with non-polynomial local free-energy terms.

(ii) Section 7 applies this technique to the coarse-grained free energy $\mathcal{F}_{CG}$ and uses it to point out the differences in the stabilities of different structures between the BDL and LP models, and in particular to explain the previously surprising robustness of decagonal structures in the BDL model.

(iii) Finally, equipped with this new understanding of the effect of the local free-energy term, we again use the density distribution method in Section 8 to generate an artificial local free-energy term that can stabilize quasicrystals with $6n$-fold symmetry, with arbitrarily large $n$, using only two length scales.

Note that the vast majority of the stable two-dimensional quasicrystals that have been discovered to date have symmetry orders no greater than 12-fold. Possible explanations for this have been suggested by Levitov (1988) and Mikhael et al. (2010). Exceptions are the octadecagonal quasicrystal discovered by Fischer et al. (2011), those found numerically (Dotera et al., 2014; Engel & Glotzer, 2014; Pattabhiraman & Dijkstra, 2017b) and the one discussed in Section 5, as well as the numerically discovered icositetragonal (24-fold) quasicrystals (Dotera et al., 2014; Engel & Glotzer, 2014).

For completeness, we should note three additional extensions of the LP and BDL models that we do not discuss here. First, in the present work we focus solely on the question of thermodynamic stability (or metastability), searching for the minimum free-energy states, without considering any actual dynamics. LP used purely relaxational dynamics, also known as Model A of Hohenberg & Halperin (1977)

$$\frac{\partial \rho}{\partial t} = -\frac{\delta \mathcal{F}_{LP}}{\delta \rho}$$
$$= \epsilon \rho - c(\nabla^2 + 1)^2 (\nabla^2 + q^2)^2 \rho + \alpha \rho^2 - \rho^3, \qquad (4)$$

starting with random initial conditions to confirm numerically that the steady states were indeed the targeted ones. This also established that quasicrystals were not as difficult to obtain as solutions of simple partial differential equations as one may have thought. Recent authors have been using conserved dynamics of the form $\partial_t \rho = D\nabla^2(\delta\mathcal{F}/\delta\rho)$, also known as Model

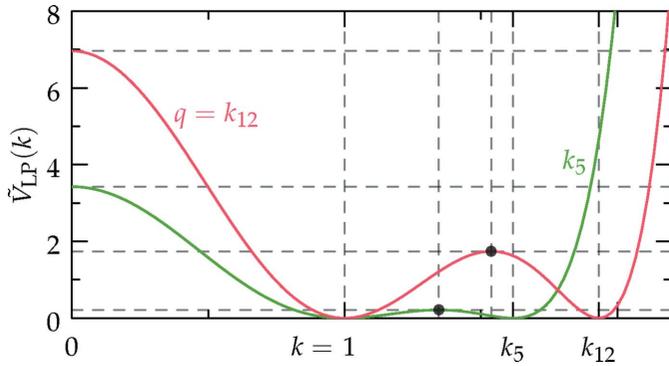

**Figure 1**
$\tilde{V}_{LP}(k)$ for $c = 1$ and $q = k_5$ and $k_{12}$: note that $\tilde{V}_{LP}(k)$ is positive for all $k$ except $k = 1$ or $q$, where it is zero. Also note that the barrier between the minima when $q = k_{12}$ is significantly larger than the barrier when $q = k_5$. This disparity is important for the study of the finite-$c$ case discussed in Section 4.

B of Hohenberg & Halperin (1977), or slight variations of Model B, known as dynamic density functional theory (DDFT) or phase-field crystal (PFC) models (Archer *et al.*, 2013; Achim *et al.*, 2014, 2015; Subramanian *et al.*, 2016).[2]

Interestingly, the PFC model of Achim *et al.* (2014) uses $\mathcal{F}_{LP}$ without the cubic term, but still genrates the same structures. At first sight, this seems to contradict LP's explanation of the stability of their quasicrystals. However, it turns out that this apparent restoration of the $\mathbb{Z}_2$ symmetry $\rho \to -\rho$ in the local term of the free energy is destroyed by the conservation of mass condition, which constrains the average density $\overline{\rho}$ to be a positive constant. This, in turn, generates an effective cubic term in the local free energy, with $\alpha_{eff} \propto \overline{\rho}$ (Barkan, 2015).

Extensions of the LP and BDL models, which we intend to pursue elsewhere, include:

(i) The application of these models in three dimensions. This has already been shown by some authors to produce stable icosahedral quasicrystals using two length scales (Subramanian *et al.*, 2016; Jiang *et al.*, 2017).[3]

(ii) The generalization to two (or more) interacting densities or order-parameter fields. The use of two coupled fields or two coupled Swift–Hohenberg equations, where each field carries one of the length scales, was considered very early on (Mermin & Troian, 1985; Sachdev & Nelson, 1985; Narasimhan & Ho, 1988; Müller, 1994) and has been resumed recently in the context of binary and ternary soft-matter systems (Dotera, 2007; Barkan, 2015; Jiang *et al.*, 2016), with new insight gained from results of the LP and BDL models.

Soft-matter quasicrystals provide rich and versatile platforms for the realization of relatively simple theoretical models as classical particles interacting *via* pre-designed pair potentials, treated either by molecular dynamics simulations or by coarse-grained mean-field theories and their Landau expansions. Such theoretical tools may be inadequate for treating atomic-scale quasicrystals, yet perfect for the fundamental study of the basic notions of the physics of quasicrystals as they appear in soft condensed matter. Armed with renewed insight from soft-matter systems and the potential to realize them directly in the laboratory, some of the outstanding fundamental questions in the field can be treated afresh, allowing one to get closer than ever to their resolution. Admittedly, our discussion here may apply only to soft condensed matter, but intriguing new analogies between soft-matter and solid-state systems continue to emerge (Lee *et al.*, 2014; Lifshitz, 2014), possibly enlarging our scope.

## 2. The Lifshitz–Petrich model and its immediate generalizations

Following the original analysis by Lifshitz & Petrich (1997), we define a scalar field $\rho(\mathbf{r})$ on the two-dimensional Cartesian plane. The Swift–Hohenberg free energy of this field (Swift & Hohenberg, 1977) can be written as

$$\mathcal{F}_{SH}[\rho(\mathbf{r})] = \int \left\{ \frac{1}{2}\left[(\nabla^2 + 1)\rho(\mathbf{r})\right]^2 + f(\rho(\mathbf{r})) \right\} d\mathbf{r}, \quad (5)$$

where $f(\phi)$ is a local contribution to the free energy, which may or may not be symmetric under the operation that replaces $\phi$ by $-\phi$ (Cross & Greenside, 2009). In what follows, we use $\rho(\mathbf{r})$ to refer to the field and $\phi$ for specific scalar values the field can take on at a given point.

The LP free energy (1) changes this to

$$\mathcal{F}_{LP} = \int \left\{ \frac{c}{2}\left[(\nabla^2 + 1)(\nabla^2 + q^2)\rho(\mathbf{r})\right]^2 + f(\rho(\mathbf{r})) \right\} d\mathbf{r}, \quad (6)$$

and sets

$$f(\phi) = -\frac{\epsilon}{2}\phi^2 - \frac{1}{3}\phi^3 + \frac{1}{4}\phi^4, \quad (7)$$

explicitly breaking the $\phi \to -\phi$ symmetry, where $c$ is assumed positive and $q$ is the ratio of the two selected length scales, which generally satisfies $1 < q \le 2$. Note that the coefficient $\alpha$ of the cubic term in equation (1) has been scaled to unity by measuring the field amplitude $\rho$ in units of $\alpha$ and consequently measuring energy in units of $\alpha^4$. The parameter $c$, which sets the length-scale selectivity of the system, and the control parameter $\epsilon$ are then measured in units of $\alpha^2$.

By substituting the Fourier transform

$$\rho(\mathbf{r}) = \int \exp(i\mathbf{k} \cdot \mathbf{r})\,\tilde{\rho}(\mathbf{k})\,d\mathbf{k}, \quad (8)$$

into the first terms of free energies like the ones in equations (5) and (6), they can be written as

$$\mathcal{F} = \int \tilde{V}(k)\,|\tilde{\rho}(\mathbf{k})|^2\,d\mathbf{k} + \int f(\rho(\mathbf{r}))\,d\mathbf{r}, \quad (9)$$

where in $\mathcal{F}_{LP}$ the function $\tilde{V}(k)$ is given by the octic polynomial,

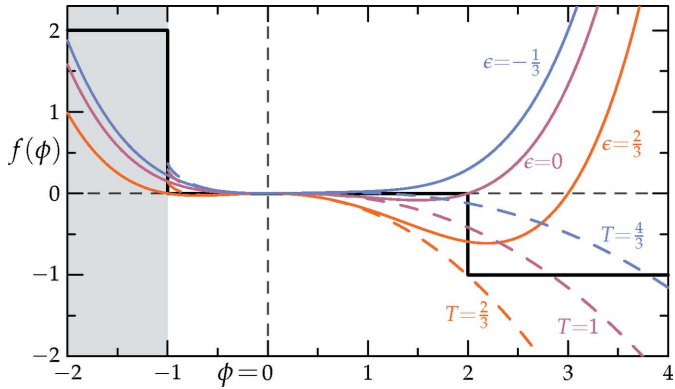

**Figure 2**
Local free-energy densities $f(\phi)$ used in this work: solid colored lines show the scaled ($\alpha = 1$) quartic local free-energy function (7) used by LP, for values of $\epsilon$ above, equal to and below the spinodal value of $\epsilon = 0$. The full logarithmic local free-energy function $f_{CG}$, in equation (49), as used by BDL, is shown with dashed lines for temperatures above, below and equal to the spinodal temperature, which is scaled to $T = 1$. The values of $\epsilon$ and $T$ are related by $T = 4/(3\epsilon + 4)$, as explained in Section 7. In order to demonstrate visually the resulting fourth-order agreement, the solid lines have been stretched horizontally by 50% while the dashed lines have been compressed vertically by a factor of 27/16. Note that the LP quartic free-energy density penetrates into the $\phi < -1$ region and diverges from the BDL logarithmic free-energy density for $\phi > 1$. Finally, the solid black line shows the free-energy density used in Section 8 to stabilize $6n$-fold quasicrystals with arbitrary $n$.

$$\tilde{V}_{LP}(k) = \frac{c}{2}\left[\left(k^2 - 1\right)\left(k^2 - q^2\right)\right]^2, \quad (10)$$

which is sketched in Fig. 1.[4] The (horizontally stretched) local quartic free-energy density (7) is plotted as the solid colored lines in Fig. 2 for different values of its single parameter $\epsilon$. The reader should note that all the position- and momentum-space integrals are implicitly normalized to give a free energy per unit area.

After rescaling, the LP model is left with only three free parameters, $q$, $c$ and $\epsilon$. Given specific values for these, we seek the configurations $\rho(\mathbf{r})$ that minimize $\mathcal{F}_{LP}$. This is easiest in the limit where $c$ is taken to infinity. Because this infinite-$c$ limit is also generally favorable for the formation of quasicrystals, we adopt it throughout this paper, with the exception of Section 4. In this limit, $\tilde{V}_{LP}(k) = 0$ if $k$ belongs to the set $\tilde{V}_0 = \{1, q\}$ and is otherwise infinite. Thus, we immediately conclude that

$$\lim_{c \to \infty} \tilde{\rho}(\mathbf{k}) = 0 \quad \text{unless} \quad k = |\mathbf{k}| \in \tilde{V}_0, \quad (11)$$

---

[4] The particular form of the function $\tilde{V}_{LP}(k)$, with its octic ultraviolet divergence, comes from duplicating the gradient term of the Swift–Hohenberg equation (5), which possesses a quartic divergence, and whose purpose is to favor modes with wavenumbers approximately equal to unity or $q$. Conveniently, after taking the infinite-$c$ limit, as we do everywhere outside of Section 4, it diverges almost everywhere, and its only remaining features are its minima at $k = 1$ and $k = q$ where $\tilde{V}_{LP}(k) = 0$. On the other hand, when viewed as the Fourier transform of a real-space interaction potential, it is incompatible with the kind of soft-matter particles we have in mind, as it introduces a strongly diverging interaction at the origin. One may, technically, tame this interaction by multiplying it with a Gaussian, as was done by Barkan *et al.* (2014).

restricting the support of $\tilde{\rho}(\mathbf{k})$ to lie entirely on two concentric circles of radii unity and $q$, centered about the origin. Given this restriction, the free energy is simply

$$\mathcal{F} = \int f(\rho(\mathbf{r}))\,d\mathbf{r}. \quad (12)$$

Upon substituting the Fourier transform (8) of the field – which for quasiperiodic density fields is supported on a countable set of wavevectors, changing the integral into a sum – this becomes

$$
\begin{aligned}
\mathcal{F} = &-\frac{\epsilon}{2}\sum_{\mathbf{k}}^{\sim} \tilde{\rho}(\mathbf{k})\,\tilde{\rho}(-\mathbf{k}) \\
&-\frac{1}{3}\sum_{\mathbf{k}_1,\mathbf{k}_2}^{\sim} \tilde{\rho}(\mathbf{k}_1)\,\tilde{\rho}(\mathbf{k}_2)\,\tilde{\rho}(-\mathbf{k}_1-\mathbf{k}_2) \\
&+\frac{1}{4}\sum_{\mathbf{k}_1,\mathbf{k}_2,\mathbf{k}_3}^{\sim} \tilde{\rho}(\mathbf{k}_1)\,\tilde{\rho}(\mathbf{k}_2)\,\tilde{\rho}(\mathbf{k}_3)\,\tilde{\rho}(-\mathbf{k}_1-\mathbf{k}_2-\mathbf{k}_3), \quad (13)
\end{aligned}
$$

where the tilde indicates the restriction that the magnitude of all the wavevectors and their sum must belong to the set $\tilde{V}_0$. The products of Fourier coefficients on wavevectors that add up to zero, appearing in this expression for the free energy, are known in crystallography as 'structure invariants'. The sums can easily be evaluated on a computer using symbolic algebra. A wise choice of $q$ can make use of the triplets, or wavevector triangles, on the second line for stabilizing the desired two-scale structures, as mentioned earlier. The benefit of adding such triplets usually comes at the cost of more quadruplets on the third line that generally increase the free energy. In Section 3, essentially by counting triplets and quadruplets, we repeat the calculation of Lifshitz & Petrich (1997) and show that this simple free energy is able to stabilize periodic square and hexagonal crystals, decagonal and dodecagonal quasicrystals, and lamellae, also called stripes.

In addition to the obvious generalization to three dimensions, the free energy in equation (9) can immediately be generalized by modifying $\tilde{V}(k)$, $f(\phi)$ or both, in a number of ways:

(i) While remaining in the infinite-$c$ limit, $\tilde{V}_{LP}(k)$ can be changed by modifying the set $\tilde{V}_0$. We show in Section 5 that doubling the cardinality of $\tilde{V}_0$, *i.e.* going from two to four concentric circles, allows us to stabilize octagonal and octadecagonal quasicrystals.

(ii) Subramanian *et al.* (2016) modify $\tilde{V}_{LP}(k)$ in a particular way in order to gain control of the relative heights of the two minima (see Fig. 1), while leaving $c$ finite.

(iii) $\tilde{V}(k)$, as indicated by its notation, can be thought of as the radial Fourier transform (also known as the Hankel transform) of an isotropic interaction potential $\mathcal{U}(r)$ of a pair of particles in real space, possibly scaled and shifted by a constant. This, along with a replacement of $f(\phi)$ by the local entropy term from $\mathcal{F}_{CG}$ in equation (3), forms the basis of the BDL model, which we consider and expand our understanding of in Section 7. A comparison of these two choices for $f(\phi)$ is shown in Fig. 2.

(iv) In Section 8 we show that an artificial yet judicious choice of $f(\phi)$ can actually stabilize $6n$-fold quasicrystals for any $n \geq 2$, with just two length scales.

## 3. Stable periodic and quasiperiodic crystals in the original LP model with exact length-scale selection

### 3.1. Notation and method of calculation

#### 3.1.1. Calculation of stability bounds.
We set $q$ equal to $k_n$ from equation (2) with $n > 3$, so that $1 < q \leq 2$, and the upper limit of 2 is obtained for $n \to \infty$. Our goal is to stabilize $N$-fold symmetric structures whose Fourier coefficients are confined to two circles of radii unity and $k_n$. Each circle is expected to contain $N$ equally separated Bragg peaks, with $N = n$ or $2n$ when $n$ is even or odd, respectively. These targeted structures are shown schematically in Figs. $3(i)$–$3(o)$ for $k_4 = 2^{1/2}$, $k_6 = 3^{1/2}$, $k_\infty = 2$, $k_5 = (1 + 5^{1/2})/2$, $k_{12} = (2 + 3^{1/2})^{1/2}$, $k_8 = (2 + 2^{1/2})^{1/2}$ and $k_{10} = [(5 + 5^{1/2})/2]^{1/2}$, respectively.

These two-scale structures are in thermodynamic competition with the uniform liquid phase $\rho(\mathbf{r}) = 0$, and with single-scale and trivial two-scale periodic structures consisting of two degenerate lamellar phases varying in their spatial scale (set by which circle the two peaks lie on[5]), four degenerate hexagonal configurations, two of which are regular and two distorted, containing two length scales, and infinitely many oblique, rectangular and square structures consisting of a sum of two cosines with an arbitrary relative orientation, whose wavevectors are taken from the set $\{1, q\}$. These competing structures are shown schematically in Figs. $3(a)$–$3(h)$. The targeted structures and the competing ones are also listed in Table 1. As is typical for these kinds of stability calculations, one cannot be certain that the list of candidate and competing structures is exhaustive. We believe it is complete, however, not only due to intuitive symmetry considerations, but also based on repeated computational simulations. As one example, the Model A dynamics of equation (4) have been applied to many realizations of random initial conditions, thereby exploring the space of likely minimum free-energy states over a wide range of model parameters.

As noted by LP, because all the candidate structures are centrosymmetric, and because there are no screw rotations in two dimensions, we may always take each of the Fourier coefficients on a given circle to be equal, and their phases may all be chosen such that they are either 0 or $\pi$, corresponding to positive and negative real values, respectively.[6] The minimization of the free energy (13) is therefore always with respect to no more than two real variables, which we denote $\tilde{\rho}$ in structures with a single scale and $\tilde{\rho}_1$ and $\tilde{\rho}_q$ in the two-scale structures. For example, the larger regular hexagonal phase has a real-space structure of

$$\rho(\mathbf{r}) = 2\tilde{\rho}\left(\cos x + \cos \frac{x + 3^{1/2}y}{2} + \cos \frac{-x + 3^{1/2}y}{2}\right). \quad (14)$$

The free energy of equation (13) then becomes a quartic function of two variables, given by

---

[5] Jiang et al. (2015) termed these 'sibling periodic crystals'.
[6] More technically, there is always a Rokhsar–Wright–Mermin gauge transformation (Rokhsar et al., 1988) that can be applied to the free-energy minimizing structure to obtain these phase values, without altering the free energy. For specific information regarding the required gauge transformation, and a definition of a screw operation in the case of quasicrystals, see, for example, Rabson et al. (1991).

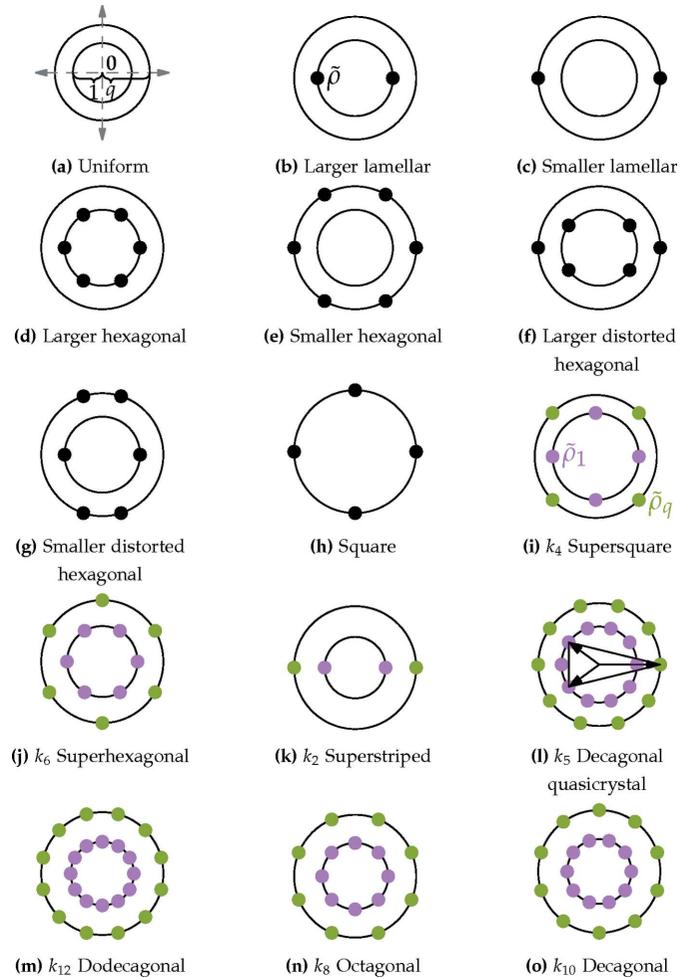

**Figure 3**
Fourier spectra of the candidate structures: structures $(a)$–$(g)$ use the arbitrary ratio $q = k_5$, while all other ratios are specified explicitly. The uniform phase $(a)$ has no Fourier modes. An example of a stabilizing triangle is included in the $k_5$ decagonal structure $(l)$.

(a) Uniform  (b) Larger lamellar  (c) Smaller lamellar
(d) Larger hexagonal  (e) Smaller hexagonal  (f) Larger distorted hexagonal
(g) Smaller distorted hexagonal  (h) Square  (i) $k_4$ Supersquare
(j) $k_6$ Superhexagonal  (k) $k_2$ Superstriped  (l) $k_5$ Decagonal quasicrystal
(m) $k_{12}$ Dodecagonal  (n) $k_8$ Octagonal  (o) $k_{10}$ Decagonal

$$\mathcal{F}\left(\tilde{\rho}_1, \tilde{\rho}_q\right) = -\frac{\epsilon}{2} \sum_{\mathbf{k}_1 + \mathbf{k}_2 = 0} \tilde{\rho}_{k_1}\tilde{\rho}_{k_2}$$
$$- \frac{1}{3} \sum_{\mathbf{k}_1 + \mathbf{k}_2 + \mathbf{k}_3 = 0} \tilde{\rho}_{k_1}\tilde{\rho}_{k_2}\tilde{\rho}_{k_3}$$
$$+ \frac{1}{4} \sum_{\mathbf{k}_1 + \mathbf{k}_2 + \mathbf{k}_3 + \mathbf{k}_4 = 0} \tilde{\rho}_{k_1}\tilde{\rho}_{k_2}\tilde{\rho}_{k_3}\tilde{\rho}_{k_4}. \quad (15)$$

for two-scale structures, where all $k_i = |\mathbf{k}_i| \in \{1, q\}$, and a similar function of the single variable $\tilde{\rho}$ for single-scale structures. Computer-assisted symbolic algebra is used to evaluate the sums, and then to minimize them with respect to $\tilde{\rho}$ or $\tilde{\rho}_1$ and $\tilde{\rho}_q$. The structure that has the lowest free energy, given the value of $\epsilon$, is the thermodynamically stable one, assuming we have not overlooked any additional competing structures.

#### 3.1.2. Calculation of metastability bounds.
The majority of work on the LP model has been focused on finding the free-energy minimizing structure for various choices of the parameters. However, as seen in the work by Barkan et al. (2014), the phase transitions between these structures exhibit significant hysteresis. As a first attempt at evaluating the meta-

**Table 1**
Stability and metastability boundaries in the infinite-$c$ LP model.

Note that the stability regions of the single-scale structures are modified by their competition with the $q$-tuned two-scale structures. The metastability regions are left unchanged, except in the $k_6$ and $k_{12}$ cases where the lower bound becomes 0. Recall that $\epsilon$ here is in units of $\alpha^2$, as in equation (7); without scaling one should multiply the quoted values by $\alpha^2$. The choices of $q = k_8$ and $k_{10}$ fail to stabilize octagonal and decagonal quasicrystals, respectively. However, they can exhibit regions of metastability, which might be observable in experiments or simulations given proper initial conditions.

| | Structure | $q$ | Stable | Metastable | Figures |
|---|---|---|---|---|---|
| | Uniform | | $\epsilon \lesssim -0.05926$ | $\epsilon \leq 0$ | 3(a) |
| One-scale periodic | Hexagonal | | $-0.05926 \lesssim \epsilon \lesssim 1.913$ | $-0.06667 \lesssim \epsilon \lesssim 5.333$ | 3(d)–3(g), 4, 15, 22 |
| | Lamellar | | $1.913 \lesssim \epsilon$ | $1.333 \lesssim \epsilon$ | 3(b), 3(c), 4, 14, 15 |
| | Square | | Unstable | Unstable | 3(h) |
| Two-scale periodic | Square | $k_4 = 2^{1/2} \simeq 1.414$ | $0.09205 \lesssim \epsilon \lesssim 0.7689$ | $0.09205 \lesssim \epsilon \lesssim 2.113$ | 3(i), 5(a) |
| | Hexagonal | $k_6 = 3^{1/2} \simeq 1.732$ | $-0.1143 \lesssim \epsilon \lesssim 2.074$ | $-0.1281 \lesssim \epsilon \lesssim 8.827$ | 3(j), 5(b) |
| | Lamellar | $k_\infty = 2$ | $-0.06904 \lesssim \epsilon$ | $-0.07760 \lesssim \epsilon$ | 3(k), 5(c) |
| Two-scale quasicrystal | Decagonal | $k_5 = (5^{1/2}+1)/2 \simeq 1.618$ | $-0.08602 \lesssim \epsilon \lesssim 0.2290$ | $-0.09677 \lesssim \epsilon$ | 1, 3(l), 6(a), 16, 21 |
| | Dodecagonal | $k_{12} = (2+3^{1/2})^{1/2} \simeq 1.932$ | $-0.1009 \lesssim \epsilon \lesssim 0.03055$ | $-0.1135 \lesssim \epsilon$ | 1, 3(m), 6(b), 16 |
| | | $k_8$, $k_{10}$ etc. | Unstable | See caption | 3(n), 3(o) |

stability bounds on $\epsilon$, we can imagine writing a structure as a linear combination of multiple components

$$\rho(\mathbf{r}) = \sum_i A_i \rho_i(\mathbf{r}), \qquad (16)$$

such as aligned lamellar, hexagonal and dodecagonal modes on a circle.

The coefficients $A_i$ are set by the local minimum of the free energy,

$$\frac{\partial \mathcal{F}[\rho(\mathbf{r})]}{\partial A_i} = 0, \qquad (17)$$

where, for a given phase, some of the $A_i$'s will be zero. These represent the potential 'directions' in which the structure can decay. The spinodal decomposition of a phase, where it is no longer metastable, occurs when that point on the free-energy landscape transitions from a local minimum to a saddle point. This occurs when the determinant of the Hessian of the free energy of this structure,

$$\frac{\partial^2 \mathcal{F}}{\partial A_i \partial A_j}, \qquad (18)$$

becomes zero.

In our calculations, we include all of the candidates as potential decay directions, but we have no proof that these are the only options, so the reader should take the metastability bounds reported below as tentative results.

### 3.2. Stability and metastability bounds in the original LP model

#### 3.2.1. Single-scale phases.
First, we consider the single-scale lamellar, hexagonal and square phases, along with the uniform liquid state. Recall, also, that this includes the two distorted hexagonal phases that have the same free energy as the two regular ones, even though, strictly speaking, they consist of two length scales. The Fourier spectra of these structures are depicted in Figs. 3(a)–3(e) and 3(h). The stability bounds are summarized in the top section of Table 1 and plotted in Fig. 4.

The uniform phase always has a free energy of

$$\mathcal{F}_{\text{UNIF}} = 0, \qquad (19)$$

and decays spinodally when $\epsilon > 0$.

For the lamellar phase, the free-energy equation (15) gives

$$\mathcal{F}_{\text{LAM}}(\tilde\rho; \epsilon) = -\epsilon\tilde\rho^2 + \frac{3}{2}\tilde\rho^4. \qquad (20)$$

Minimizing $\mathcal{F}_{\text{LAM}}$ over $\tilde\rho$ shows that, for $\epsilon < 0$, $\tilde\rho = 0$, thus giving a uniform phase. The lamellar phase therefore only exists for positive $\epsilon$, wherein

$$\mathcal{F}_{\text{LAM}}(\epsilon) = -\frac{\epsilon^2}{6}. \qquad (21)$$

The free energy of the hexagonal phase is given by

$$\mathcal{F}_{\text{HEX}}(\tilde\rho; \epsilon) = -3\epsilon\tilde\rho^2 - 4\tilde\rho^3 + \frac{45}{2}\tilde\rho^4. \qquad (22)$$

It only exists for $\epsilon \geq -1/15$, below which the nontrivial minima of equation (22) are complex, and so the only possible state is the uniform one with $\tilde{\rho} = 0$. Above this point – a saddle node in the context of dynamical bifurcation theory – we have the nontrivial minimum

$$\mathcal{F}_{\text{HEX}}(\epsilon) = \frac{-675\epsilon^2 - 8(15\epsilon + 1)^{3/2} - 180\epsilon - 8}{6750}. \quad (23)$$

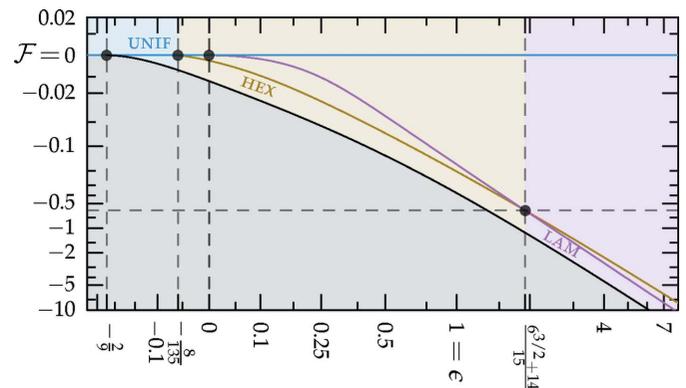

**Figure 4**
Free energies of the single-scale periodic structures in the infinite-$c$ LP model: as $\epsilon$ is increased, the uniform liquid is the equilibrium phase until it reaches $-8/135$, at which point a first-order transition to the hexagonal structure occurs. This persists until $\epsilon$ reaches $(6^{3/2} + 14)/15$ where the lamellar structure becomes the equilibrium phase. The gray region corresponds to the forbidden zone below the lower bound, calculated in Section 6.6.

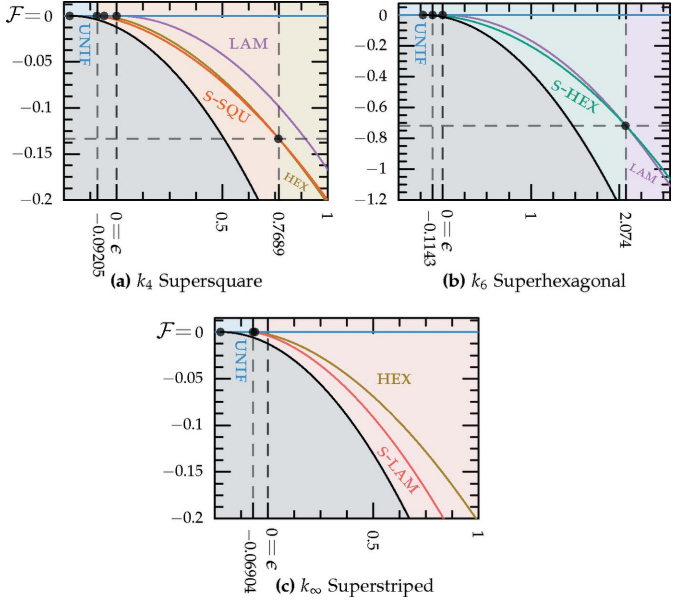

**Figure 5**
Free energies of the two-scale periodic structures in the infinite-$c$ LP model: in panel ($a$), with $q = k_4$, the square superstructure dominates when $-0.09205 \lesssim \epsilon \lesssim 0.7689$. In panel ($b$), with $q = k_6$, the hexagonal superstructure dominates when $-0.1143 \lesssim \epsilon \lesssim 2.074$. Finally, in panel ($c$), with $q = k_\infty$, the lamellar superstructure dominates when $\epsilon \gtrsim -0.06904$. The two-scale superstructure always has a lower free energy than its single-scale analogue. The gray region is the same as in Fig. 4.

For generic $q$, as $\epsilon$ is increased, the system first undergoes a first-order transition from the uniform phase to the hexagonal phase at $\epsilon = -8/135 \simeq -0.05926$ and then to the lamellar phase at $\epsilon = (6^{3/2} + 14)/15 \simeq 1.913$. At the second transition, $\mathcal{F} = -[84(6^{1/2}) + 206]/675$. This behavior is shown in Fig. 4.

The hexagonal phase is metastable when $-0.06667 \simeq -1/15 \leq \epsilon \leq 16/3 \simeq 5.333$, and the lamellar phase is metastable for all $\epsilon \geq 4/3 \simeq 1.333$.

The single-scale square structure in Fig. 3($h$) and its infinitely many degenerate oblique, rectangular and square structures, consisting of the sum of two cosines with an arbitrary relative orientation, all have a free energy of

$$\mathcal{F}_{\text{SQU}}(\tilde{\rho}; \epsilon) = -2\epsilon\tilde{\rho}^2 + 9\tilde{\rho}^4, \quad (24)$$

which leads to a minimized free energy of

$$\mathcal{F}_{\text{SQU}}(\epsilon) = -\frac{\epsilon^2}{9}. \quad (25)$$

Because the square structure has additional quadruplets compared with the lamellar phase (21) without any compensating triplets, its free energy is always higher, and it is therefore never in thermodynamic equilibrium.

### 3.2.2. Two-scale periodic phases.
Next, we consider the two-scale square, hexagonal and striped superstructures, for $q$

$= k_4$, $k_6$ and $k_\infty$, respectively. Their Fourier spectra are shown in Figs. 3($i$)–3($k$).[7] For these structures there are no simple expressions for the minimized $\tilde{\rho}_1$ and $\tilde{\rho}_q$ values, which are generally unequal, nor for their minimized free energies and critical $\epsilon$ values. Thus, we provide only numerical results for the stability bounds in the middle section of Table 1 and plot these bounds in Fig. 5.

The free energies from which these bounds are obtained are given by

$$\mathcal{F}_{\text{S-SQU}}(\tilde{\rho}_1, \tilde{\rho}_q; \epsilon) = \mathcal{F}_{\text{SQU}}(\tilde{\rho}_1; \epsilon) + \mathcal{F}_{\text{SQU}}(\tilde{\rho}_q; \epsilon) + 4\tilde{\rho}_1^2\tilde{\rho}_q(-2 + 9\tilde{\rho}_q), \quad (26a)$$

$$\mathcal{F}_{\text{S-HEX}}(\tilde{\rho}_1, \tilde{\rho}_q; \epsilon) = \mathcal{F}_{\text{HEX}}(\tilde{\rho}_1; \epsilon) + \mathcal{F}_{\text{HEX}}(\tilde{\rho}_q; \epsilon) + 6\tilde{\rho}_1^2\tilde{\rho}_q(-2 + 6\tilde{\rho}_1 + 15\tilde{\rho}_q), \quad (26b)$$

$$\mathcal{F}_{\text{S-LAM}}(\tilde{\rho}_1, \tilde{\rho}_q; \epsilon) = \mathcal{F}_{\text{LAM}}(\tilde{\rho}_1; \epsilon) + \mathcal{F}_{\text{LAM}}(\tilde{\rho}_q; \epsilon) + 2\tilde{\rho}_1^2\tilde{\rho}_q(-1 + 3\tilde{\rho}_q), \quad (26c)$$

where $\mathcal{F}_{\text{SQU}}(\tilde{\rho}_q; \epsilon)$, $\mathcal{F}_{\text{LAM}}(\tilde{\rho}_q; \epsilon)$ and $\mathcal{F}_{\text{HEX}}(\tilde{\rho}_q; \epsilon)$ are given in equations (24), (22) and (20), respectively.

### 3.2.3. Two-scale quasiperiodic phases.
Finally, we consider the two-scale quasicrystals with $q = k_5$, $k_{12}$, $k_8$ and $k_{10}$, whose Fourier spectra are shown in Figs. 3($l$)–3($o$), respectively. We find that the $k_8$ and $k_{10}$ structures are never global minima of the free energy and are therefore unstable. We do not give the detailed calculation of their free energies here, and only note that they may exhibit regions of metastability. Thus, one could potentially observe them in experiment or simulation given proper initial conditions. The stability bounds for the $k_5$ decagonal quasicrystal and the $k_{12}$ dodecagonal quasicrystal are included in the latter half of Table 1 and plotted in Fig. 6. The stability bounds reported here should be taken in place of the original bounds reported by Lifshitz & Petrich (1997), as they missed the existence of a stable decagonal quasicrystal with $q = k_5$ rather than $k_{10}$ and miscalculated the stability boundaries of the dodecagonal one.[8]

The simplest expressions for the upper stability bounds, for both of these phases, involve roots of quintic polynomials that are provided below for the first time. The metastability bounds for all of the structures studied in this section are reported here for the first time as well. Because all of the results are summarized in Table 1, readers who are not interested in the detailed calculation itself may skip to the next section.

The free energy of the $k_5$ decagonal phase is given by

$$\mathcal{F}_{\text{DEC}}(\tilde{\rho}_1, \tilde{\rho}_q; \epsilon) = -5\epsilon(\tilde{\rho}_1^2 + \tilde{\rho}_q^2) - 20(\tilde{\rho}_1^2\tilde{\rho}_q + \tilde{\rho}_1\tilde{\rho}_q^2) \\ + \frac{135}{2}(\tilde{\rho}_1^4 + \tilde{\rho}_q^4) + 60(\tilde{\rho}_1^3\tilde{\rho}_q + \tilde{\rho}_1\tilde{\rho}_q^3) \\ + 210\tilde{\rho}_1^2\tilde{\rho}_q^2. \quad (27)$$

---

[7] The single-scale lamellar, square and hexagonal structures, and the two-scale square and hexagonal superstructures, correspond to the $A_1$, $A_1 \times A_1$, $A_2$, $B_2$ and $G_2$ Lie algebra root systems, respectively.
[8] These discrepancies led to some confusion in the subsequent literature [see, e.g., p. 3 and footnote 32 of Barkan et al. (2011), p. 2 and footnote 26 of Barkan et al. (2014), and pp. 7 and 9 of Jiang et al. (2015)].

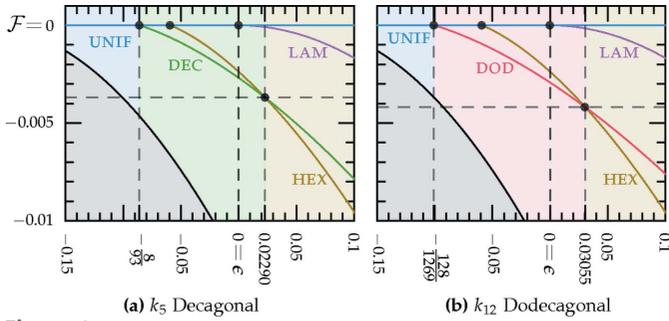

$\mathcal{F} = 0$

**(a)** $k_5$ Decagonal  **(b)** $k_{12}$ Dodecagonal

**Figure 6**
Free energies of the quasiperiodic structures in the infinite-$c$ LP model: in panel (a), with $q = k_5$, the decagonal structure dominates when $-8/93 \le \epsilon \lesssim 0.02290$. In panel (b), with $q = k_{12}$, the dodecagonal structure dominates when $-128/1269 \le \epsilon \lesssim 0.03055$. The gray region is the same as in Fig. 4.

It exists only when $\epsilon \ge -3/31 \simeq -0.09677$. For $-3/31 \le \epsilon \lesssim 0.5245$, $\tilde\rho_1 = \tilde\rho_q$ and

$$\mathcal{F}_{\text{DEC}}(\epsilon)$$
$$= \frac{-15 \times 31^2 \epsilon^2 - 40(3^{1/2})(31\epsilon + 3)^{3/2} - 180 \times 31\epsilon - 360}{9 \times 31^3}. \tag{28}$$

Above this approximate upper bound, for which there is no simple expression, the free energy continues to decrease, but $\tilde\rho_1$ no longer equals $\tilde\rho_q$. The free energy at the transition is approximately $-0.04889$.

The free energy of the dodecagonal phase is given by

$$\begin{aligned}\mathcal{F}_{\text{DOD}}(\tilde\rho_1, \tilde\rho_q; \epsilon) = &-6\epsilon(\tilde\rho_1^2 + \tilde\rho_q^2) - 8(\tilde\rho_1^3 + \tilde\rho_q^3) \\ &- 24(\tilde\rho_1^2\tilde\rho_q + \tilde\rho_1\tilde\rho_q^2) + 99(\tilde\rho_1^4 + \tilde\rho_q^4) \\ &+ 144(\tilde\rho_1^3\tilde\rho_q + \tilde\rho_1\tilde\rho_q^3) + 360\tilde\rho_1^2\tilde\rho_q^2.\end{aligned} \tag{29}$$

It exists only for $\epsilon \ge -16/141 \simeq -0.1135$, where

$$\mathcal{F}_{\text{DOD}}(\epsilon)$$
$$= \frac{54 \times 47^2 \epsilon^2 - 2^6(141\epsilon + 16)^{3/2} - 9 \times 2^7 \times 47\epsilon - 2^{12}}{27 \times 47^3} \tag{30a}$$

for $-16/141 \le \epsilon \le -208/867 \simeq 0.2399$, and

$$\mathcal{F}_{\text{DOD}}(\epsilon)$$
$$= \frac{-81 \times 67^2 \epsilon^2 - 8(3^{1/2})(67\epsilon + 75)^{3/2} - 180 \times 67\epsilon - 9000}{9 \times 67^3} \tag{30b}$$

for $208/867 \le \epsilon$. In the first case, $\tilde\rho_1 = \tilde\rho_q$, but in the second, $\tilde\rho_1 \ne \tilde\rho_q$. The free energy at the transition between these two free-energy minima is $-2^9 \times 73/(3^3 \times 17^4)$.

It is interesting to note that the free energies of both the decagonal and dodecagonal quasicrystals, in the infinite-$c$ limit, have an additional, accidental, $\mathbb{Z}_2$ symmetry associated with the exchange of $\tilde\rho_1$ and $\tilde\rho_q$. In both cases, when minimizing the free energy with respect to these amplitudes there is one solution branch that maintains this symmetry with $\tilde\rho_1 = \tilde\rho_q$ and a second branch where the $\mathbb{Z}_2$ symmetry is sponta-neously broken and $\tilde\rho_1 \ne \tilde\rho_q$. Yet, as it turns out, in the decagonal quasicrystal this symmetry-breaking transition is first order, while in the dodecagonal case it is a continuous phase transition.

We compare the free energies of these quasicrystal phases, given the requisite value of $q$, with those of the uniform and hexagonal phases in Fig. 6, which shows that the decagonal phase is stable for $-0.08602 \simeq -8/93 \le \epsilon \lesssim 0.02290$. The upper bound is given by the second real root of the quintic

$$3^5 \times 25 \times 31^2 \times 43^4 \, \epsilon^5 - 16 \times 3^5 \times 5 \times 31 \times 43 \times 103\,703 \, \epsilon^4$$
$$- 64 \times 81 \times 4\,938\,418\,073 \, \epsilon^3 - 2^{10} \times 27 \times 5 \times 11\,798\,281 \, \epsilon^2$$
$$- 2^9 \times 3 \times 5 \times 1\,046\,081 \, \epsilon + 2^{12} \times 5 \times 67\,303 = 0, \tag{31a}$$

and the free energy at the transition is approximately $-3.694 \times 10^{-3}$.

The dodecagonal phase is stable for $-0.1009 \simeq -128/1269 \le \epsilon \lesssim 0.03055$. This upper bound is given by the second real root of

$$3^{19} \times 25 \times 47^2 \, \epsilon^5 - 16 \times 3^{10} \times 5 \times 121 \times 47 \times 7757 \, \epsilon^4$$
$$- 64 \times 27 \times 7 \times 4\,093\,625\,687 \, \epsilon^3 - 2^{10} \times 27 \times 7 \times 16\,491\,709 \, \epsilon^2$$
$$+ 2^{12} \times 9 \times 11 \times 27\,953 \, \epsilon + 2^{19} \times 8059 = 0, \tag{31b}$$

and the free energy at the transition is approximately $-4.180 \times 10^{-3}$.

The decagonal phase with $\tilde\rho_1 = \tilde\rho_q$ is metastable when $-0.09677 \simeq -3/31 \le \epsilon \le 763/972 \simeq 0.7850$. The decagonal phase with $\tilde\rho_1 \ne \tilde\rho_q$ appears to be metastable for all $\epsilon \gtrsim -0.01493$.

The dodecagonal phase with $\tilde\rho_1 = \tilde\rho_q$ is metastable for $-0.1135 \simeq -16/141 \le \epsilon \le 208/867 \simeq 0.2399$. The phase with $\tilde\rho_1 \ne \tilde\rho_q$ appears to be metastable for all $\epsilon \ge 208/867$. Additionally, when $q = k_{12}$, the lower bound of the hexagonal metastability region is increased to zero.

## 4. Relaxing the requirement of exact length-scale selection

### 4.1. The two-ring approximation

Despite the convenience of taking the infinite-$c$ limit in the analysis of $\mathcal{F}_{\text{LP}}$, as given by equations (6) and (7), realistic systems can never fully extinguish all unwanted Fourier modes. It is therefore important to examine the LP model with finite-$c$ values. Evaluating the finite-$c$ LP free energy with quantitative precision requires an approach like the projection method of Jiang & Zhang (2014), which has been applied to the LP model (Jiang *et al.*, 2015). However, one can obtain a fair understanding of the role of length-scale selectivity by employing a simple 'two-ring' approximation.

We restrict $\tilde\rho(\mathbf{k})$ to lie within two rings centered about the origin. This is in contrast with allowing $\tilde\rho(\mathbf{r})$ to vary freely, as most numerical simulations do (Lifshitz & Petrich, 1997; Barkan *et al.*, 2011), or restricting $\tilde\rho(\mathbf{k})$ to some subset of a two-dimensional or higher-dimensional lattice, as in the projection method (Jiang *et al.*, 2015). This two-ring approximation compromises the numerical accuracy of our results, but what we lose in quantitative correctness we make up for in the simplicity with which we demonstrate the qualitative importance of length-scale selectivity in stabilizing quasicrystals *via* two preferred scales.

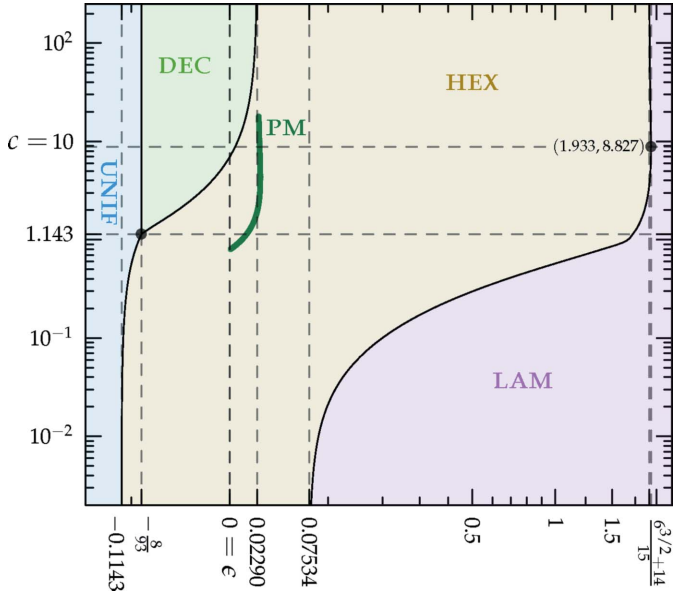

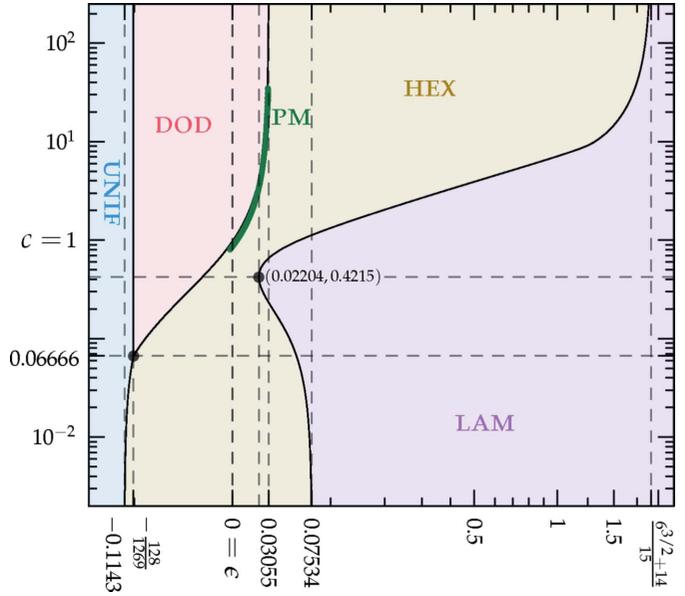

**Figure 7**
Phase diagram of the LP model with $q = k_S$ in the two-ring approximation: the thick dark-green hexagonal–decagonal boundary, which was calculated using the projection method by Jiang *et al.* (2015), indicates that this phase diagram should be considered only qualitatively. Nevertheless, note the expected uniform–hexagonal–decagonal triple point, and the possibility that the lamellar–hexagonal coexistence line has a maximum $\epsilon$ for intermediate $c$ before it reaches its expected value in the limit of infinite $c$.

**Figure 8**
Phase diagram of the LP model with $q = k_{12}$ in the two-ring approximation: in the dodecagonal case, the thick dark-green hexagonal–dodecagonal boundary, which was calculated using the projection method of Jiang *et al.* (2015), shows very good agreement with the results of the two-ring approximation. Note the uniform–hexagonal–dodecagonal triple point and the lamellar–hexagonal coexistence curve exhibiting a turning point at intermediate $c$ with a minimum value of $\epsilon$.

With this approximation, the target decagonal or dodecagonal quasicrystals have their Fourier amplitudes on exact circles of radii unity and $q$ as before, and so their free energies are unchanged. The competing lamellar and hexagonal phases are rescaled by a factor $v$ so as to position both their first and second harmonics to fit, as well as possible, within two finite-width rings near the minima of $\tilde{V}_{LP}(k)$ from equation (10), which is plotted with $c = 1$ in Fig. 1. This lowers the free energies of these phases by adding triplets to the calculation of the local contribution to the free energy in equation (9), as with the superstructures in Section 3.2.2, but comes at a cost in the integral over $\tilde{V}_{LP}(k)$,

$$n\big[\tilde{V}(v)\tilde{\rho}_v^2 + \tilde{V}(kv)\tilde{\rho}_{kv}^2\big], \tag{32}$$

where $n$ and $k$ are both 2 for the lamellar phase, and 6 and $3^{1/2}$, respectively, for the hexagonal one. Altogether, this gives

$$\mathcal{F}_{LAM}(\tilde{\rho}_v, \tilde{\rho}_{kv}, v; \epsilon) = \big[2\tilde{V}(v) - \epsilon\big]\tilde{\rho}_v^2 + \big[2\tilde{V}(2v) - \epsilon\big]\tilde{\rho}_{2v}^2$$
$$- 2\tilde{\rho}_v^2\tilde{\rho}_{2v} + \frac{3}{2}\big(\tilde{\rho}_v^4 + \tilde{\rho}_{2v}^4\big) + 6\tilde{\rho}_v^2\tilde{\rho}_{2v}^2, \tag{33a}$$

and

$$\mathcal{F}_{HEX}(\tilde{\rho}_v, \tilde{\rho}_{kv}, v; \epsilon) = \big[6\tilde{V}(v) - 3\epsilon\big]\tilde{\rho}_v^2 + \big[6\tilde{V}(3^{1/2}v) - 3\epsilon\big]\tilde{\rho}_{3^{1/2}v}^2$$
$$- 4\big(\tilde{\rho}_v^3 + \tilde{\rho}_{3^{1/2}v}^3\big) - 12\tilde{\rho}_v^2\tilde{\rho}_{3^{1/2}v}$$
$$+ \frac{45}{2}\big(\tilde{\rho}_v^4 + \tilde{\rho}_{3^{1/2}v}^4\big) + 36\tilde{\rho}_v^3\tilde{\rho}_{3^{1/2}v}$$
$$+ 90\tilde{\rho}_v^2\tilde{\rho}_{3^{1/2}v}^2. \tag{33b}$$

These are minimized numerically over $\tilde{\rho}_v$, $\tilde{\rho}_{kv}$ and $v$ for each value of $c$ and $\epsilon$ and compared with the free energies of the decagonal or dodecagonal quasicrystals, so that the minimum free-energy phase can be identified.

### 4.2. c-Dependent phase diagrams for decagonal and dodecagonal quasicrystals

The $c$-dependent phase diagrams, calculated using the two-ring approximation, are shown in Figs. 7 and 8, for the decagonal and dodecagonal quasicrystals, respectively. In both cases, one clearly observes that length selectivity, as parameterized by $c$, is a key factor contributing to quasicrystal stability. As $c$ is decreased, the competing phases change from a solution where $v = 1$ and $\tilde{\rho}_{kv} = 0$ to one where $v$ is shifted and both $\tilde{\rho}_v$ and $\tilde{\rho}_{kv}$ are nonzero, in order to take advantage of both minima of $\tilde{V}_{LP}(k)$ in an optimal way. This causes the upper bound of $\epsilon$ for quasicrystal stability to constrict with decreasing $c$ until the quasicrystalline phase vanishes. This vanishing occurs at a uniform–hexagonal–quasicrystal triple point. Below the triple point, the critical $\epsilon$ for the uniform–hexagonal transition continues to decrease towards the value $\sim -0.1143$ at zero $c$, as calculated earlier for the two-scale hexagonal phase in Section 3.2.2.

The portion of the hexagonal to quasicrystal phase boundary, calculated by Jiang *et al.* (2015) using the more accurate projection method,[9] is shown on both phase diagrams using thick dark-green lines. These lines indicate that, while both approximate phase diagrams qualitatively agree with the projection-method calculation, only the dodecagonal phase diagram agrees with it quantitatively. This is because, at the relevant $c$ values, the free-energy barrier between the two minima of $\tilde{V}_{LP}(k)$ in the decagonal ($q = k_5$) case, shown in Fig. 1, is sufficiently low that many higher-harmonic peaks appear in this region and stabilize the decagonal phase relative to the hexagonal one, which has no additional Fourier peaks there. This enlarges the stability region of the decagonal phase compared with what we calculated by restricting its Fourier coefficients to two exact circles. On the other hand, the free-energy barrier between the two minima of $\tilde{V}_{LP}(k)$ in the dodecagonal ($q = k_{12}$) case is much steeper, preventing additional rings from forming. This leads to the two-ring approximation and its predictions for the position of the triple point and the precise shapes of the phase-boundary curves, being quantitatively reasonable when $q = k_{12}$ but only qualitatively valid when $q = k_5$.

## 5. Four-scale octagonal and octadecagonal quasicrystals

Lifshitz & Petrich (1997) speculated that, with more than just two length scales, the LP model could stabilize quasicrystals with higher orders of symmetry than the dodecagonal structures they obtained, such as 18- or 24-fold. In the meantime, such structures have been observed experimentally (Fischer *et al.*, 2011) and in simulations (Engel & Glotzer, 2014; Dotera *et al.*, 2014; Pattabhiraman & Dijkstra, 2017b). In addition, Arbell & Fineberg (2002) discovered patterns with 8-fold symmetry in Faraday wave experiments using three driving frequencies. Here, we study the infinite-$c$ LP model with four length scales, by modifying $\tilde{V}_0$ in equation (11) from $\{1, q\}$ to $\{q_1, 1, q_2, q_3\}$. We consider two cases: (i) octagonal quasicrystals, with $q_1 = k_{8/3} = 2\cos(3\pi/8) = (2 - 2^{1/2})^{1/2} \simeq 0.7654$, $q_2 = k_4$ and $q_3 = k_8 = (2 + 2^{1/2})^{1/2} \simeq 1.848$; and (ii) octadecagonal quasicrystals, with $q_1 = k_{18/7} \simeq 0.6840$, $q_2 = k_{18/5} \simeq 1.286$ and $q_3 = k_{18} \simeq 1.970$. The anticipated diffraction spectra of these two structures are shown in Fig. 9.

With more than two length scales, one must carefully check for competing structures, additional to the single-scale phases in Section 3.2.1. For the octagonal case, we must consider the two-scale square superstructure shown in Fig. 3(i), allowed by the fact that $q_2 = k_4$.

For the octadecagonal case, additional competing structures stem from the fact that $q_1 + q_2 = q_3$. This allows for the 'modified' lamellar and hexagonal candidates shown in Fig. 10. These have free energies of

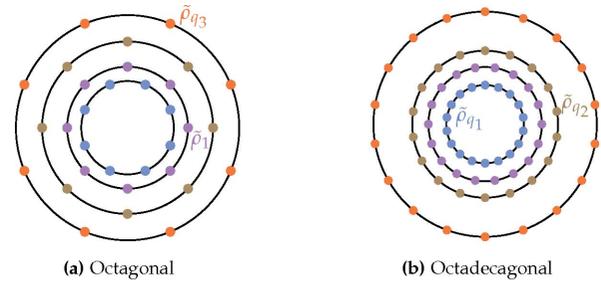

**Figure 9**
Fourier spectra of the four-scale quasicrystals: (*a*) for the octagonal quasicrystal, the radii of the circles from inside to outside are $k_{8/3}$, 1, $k_4$ and $k_8$; (*b*) for the octadecagonal quasicrystal, the radii are $k_{18/7}$, 1, $k_{18/5}$ and $k_{18}$.

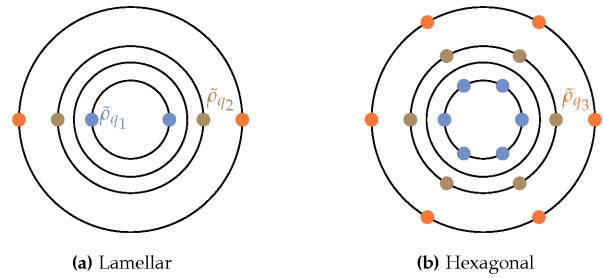

**Figure 10**
Fourier spectra of additional structures competing with the four-scale octadecagonal quasicrystal: the radii of the circles are the ones listed for Fig. 9(*b*).

$$\mathcal{F}_{\text{LAM}}(\{\tilde{\rho}\}; \epsilon) = -\epsilon \tilde{\rho}_{q_\Sigma}^2 - 4\tilde{\rho}_{q_\Pi} - \frac{3}{2}\tilde{\rho}_{q_\Sigma}^4 + 3(\tilde{\rho}_{q_\Sigma}^2)^2, \quad (34a)$$

and

$$\begin{aligned}\mathcal{F}_{\text{HEX}^*}(\{\tilde{\rho}\}; \epsilon) = {}&{-}3\epsilon \tilde{\rho}_{q_\Sigma}^2 - 4\tilde{\rho}_{q_\Sigma}^3 - 12\tilde{\rho}_{q_\Pi} \\ &- \frac{9}{2}\tilde{\rho}_{q_\Sigma}^4 + 27(\tilde{\rho}_{q_\Sigma}^2)^2 + 36\tilde{\rho}_{q_\Pi}\tilde{\rho}_{q_\Sigma},\end{aligned} \quad (34b)$$

where $\tilde{\rho}_{q_\Sigma}^n = \sum_{i=1}^3 \tilde{\rho}_{q_i}^n$ and $\tilde{\rho}_{q_\Pi} = \prod_{i=1}^3 \tilde{\rho}_{q_i}$. Minimizing these equations in the relevant $\epsilon$ range shows that the coincidental lamellar phase has $\tilde{\rho}_{q_1} = \tilde{\rho}_{q_2} = \tilde{\rho}_{q_3}$ and a free energy degenerate with the single-scale hexagonal phase. The coincidental hexagonal phase has only one ring of active modes when its free energy is minimized and so does not have its free energy reduced relative to the single-scale hexagonal structure. Thus, we can continue treating the usual single-scale hexagonal phase as the only candidate competing with the octadecagonal quasicrystal for the relevant values of $\epsilon$.

Applying equation (15) to the octagonal structure in Fig. 9(*a*) gives a free energy of

$$\begin{aligned}&\mathcal{F}_{\text{OCT}}(\{\tilde{\rho}\}; \epsilon) \\ &= -4\epsilon \tilde{\rho}_{q_\Sigma}^2 - 16\Big[\tilde{\rho}_1^2 \tilde{\rho}_{q_\Sigma} + \tilde{\rho}_{q_2}(\tilde{\rho}_{q_1} + \tilde{\rho}_{q_3})^2\Big] \\ &\quad + 6\Big(\tilde{\rho}_1^2 \Big\{-\tilde{\rho}_1^4 + 4\Big[2\tilde{\rho}_{q_\Sigma}^2 + 4(\tilde{\rho}_{q_\Sigma})^2 - 2\tilde{\rho}_{q_1}\tilde{\rho}_{q_3} - \tilde{\rho}_{q_2}^2\Big]\Big\} \\ &\quad - 5\tilde{\rho}_{q_\Sigma}^4 + 12(\tilde{\rho}_{q_\Sigma})^2 + 8\tilde{\rho}_{q_1}\tilde{\rho}_{q_3}(\tilde{\rho}_{q_\Sigma}^2 + 3\tilde{\rho}_{q_2}^2)\Big), \quad (35a)\end{aligned}$$

[9] We thank the authors for graciously sharing their raw data with us.

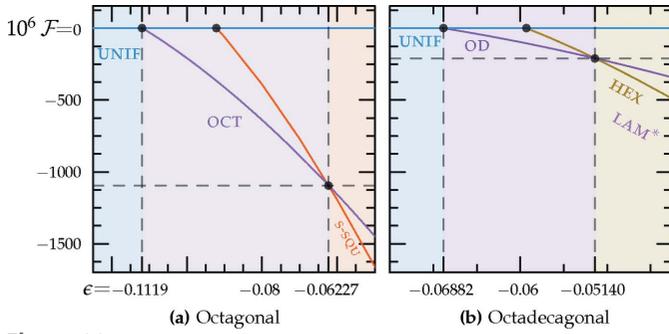

**Figure 11**
Free energies of four-scale quasicrystals in the infinite-$c$ LP model: ($a$) the octagonal structure is the equilibrium phase when $-0.1119 \lesssim \epsilon \lesssim -0.06227$; ($b$) the octadecagonal structure is the equilibrium phase when $-0.06882 \lesssim \epsilon \lesssim -0.05140$.

where $\tilde{\rho}_{q_\Sigma}^n = \tilde{\rho}_1^n + \tilde{\rho}_{q_\Sigma}^n$. While it is lengthy, it is not difficult for a computer to minimize this quartic numerically over the four $\tilde{\rho}$'s for each value of $\epsilon$. Interestingly, despite the quartic lacking $\tilde{\rho}_{q_1} \leftrightarrow \tilde{\rho}_{q_2}$ and $\tilde{\rho}_{q_2} \leftrightarrow \tilde{\rho}_{q_3}$ symmetry, the minima in the relevant small $\epsilon$ range all satisfy $\tilde{\rho}_{q_1} = \tilde{\rho}_{q_2} = \tilde{\rho}_{q_3}$.

As shown in Fig. 11($a$), the octagonal quasicrystal is expected to be thermodynamically stable when $-0.1119 \lesssim \epsilon \lesssim -0.06227$. In this range, the optimized $\tilde{\rho}_1$ is larger than the equal $\tilde{\rho}_q$'s by a factor varying between roughly 2.1 and 1.7. The free energy at the transition to the two-scale supersquare phase is approximately $-1.094 \times 10^{-3}$. A finite section of this quasicrystal is shown in Fig. 12.

The free energy of the octadecagonal structure is

$$\mathcal{F}_{OD}(\{\tilde{\rho}\}; \epsilon) = -9\epsilon\tilde{\rho}_\Sigma^2 - 12\tilde{\rho}_\Sigma^3$$
$$- 36(\tilde{\rho}_1^2\tilde{\rho}_{q_\Sigma} + \tilde{\rho}_{q_\Pi} + \tilde{\rho}_{q_1}^2\tilde{\rho}_{q_2} + \tilde{\rho}_{q_1}\tilde{\rho}_{q_3}^2 + \tilde{\rho}_{q_2}^2\tilde{\rho}_{q_3})$$
$$+ \frac{459}{2}\tilde{\rho}_\Sigma^4 + 54\tilde{\rho}_1^2\left[4\tilde{\rho}_1\tilde{\rho}_{q_\Sigma} + 5(\tilde{\rho}_{q_\Sigma})^2 + 8\tilde{\rho}_{q_\Sigma}^2\right]$$
$$+ 540\tilde{\rho}_{q_\Pi}\tilde{\rho}_{q_\Sigma} + 162(\tilde{\rho}_{q_1}^3\tilde{\rho}_{q_2} + \tilde{\rho}_{q_2}^3\tilde{\rho}_{q_3})$$
$$+ 108(\tilde{\rho}_{q_1}^3\tilde{\rho}_{q_3} + \tilde{\rho}_{q_1}\tilde{\rho}_{q_3}^3 + \tilde{\rho}_{q_2}\tilde{\rho}_{q_3}^3)$$
$$+ 675\tilde{\rho}_{q_1}^2(\tilde{\rho}_{q_2}^2 + \tilde{\rho}_{q_3}^2) + 189\tilde{\rho}_{q_1}\tilde{\rho}_{q_3}^3 + 702\tilde{\rho}_{q_2}^2\tilde{\rho}_{q_3}^2.$$
$$(35b)$$

As shown in Fig. 11($b$), the octadecagonal quasicrystal is expected to be stable when $-0.06882 \lesssim \epsilon \lesssim -0.05140$. In this range, the optimized $\tilde{\rho}_1$ is larger than the $\tilde{\rho}_q$'s by a factor varying between roughly 2.2 and 2.6. The $\tilde{\rho}_q$'s are almost, but not exactly, equal, varying by about 1%. The free energy at the transition to the hexagonal phase is approximately $-2.085 \times 10^{-4}$. A finite section of this quasicrystal is shown in Fig. 13.

## 6. The density distribution method

### 6.1. The notion of a density distribution and its quantum density of states analogy

When the local free-energy function $f(\phi)$ is a finite polynomial such as equation (7) and the structure $\rho(\mathbf{r})$ consists of a finite number of harmonic components, the free energy of a given candidate configuration can be evaluated using the approach of equation (15). However, this is not the case when the local free-energy function is non-polynomial. We describe here an alternative technique, which we call the 'density distribution method', that not only allows us to evaluate such free energies, but also provides a novel way of understanding the stability of the various periodic and quasiperiodic phases. This understanding is applied in Section 7 to calculate the free energy (3) of the candidates under the BDL model and explain the surprising stability of certain decagonal quasicrystals, and in Section 8 to aid in the artificial design of a new local-energy function $f(\phi)$ which stabilizes arbitrarily high-order quasicrystals with only two length scales.

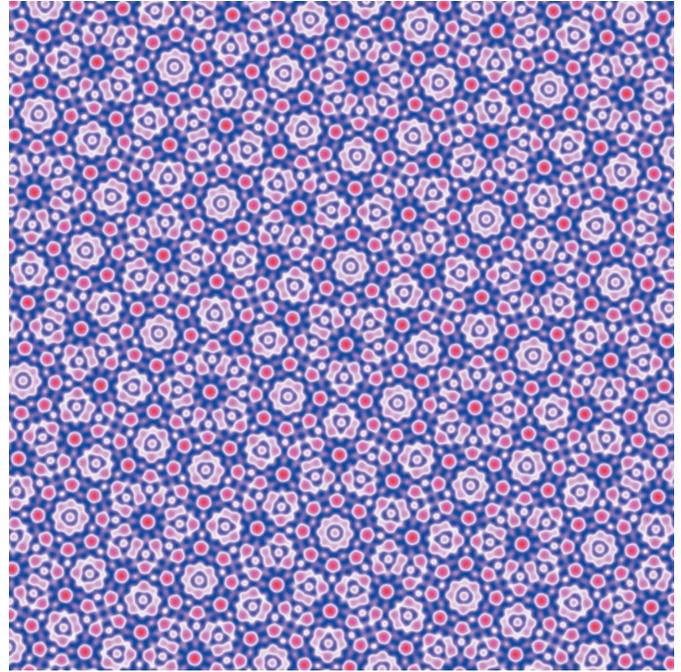

**Figure 12**
Predicted four-scale octagonal quasicrystal: blue and red shades correspond to negative field values $\phi \gtrsim -0.2306$ and positive values $\phi \lesssim 1.117$, respectively. This quasicrystal has $\epsilon = -0.09$. At this $\epsilon$, the minimization of the quartic energy in equation (35$a$) gives $\tilde{\rho}_1 \simeq 0.05427$ and $\tilde{\rho}_{q_1} = \tilde{\rho}_{q_2} = \tilde{\rho}_{q_3} \simeq 0.02856$.

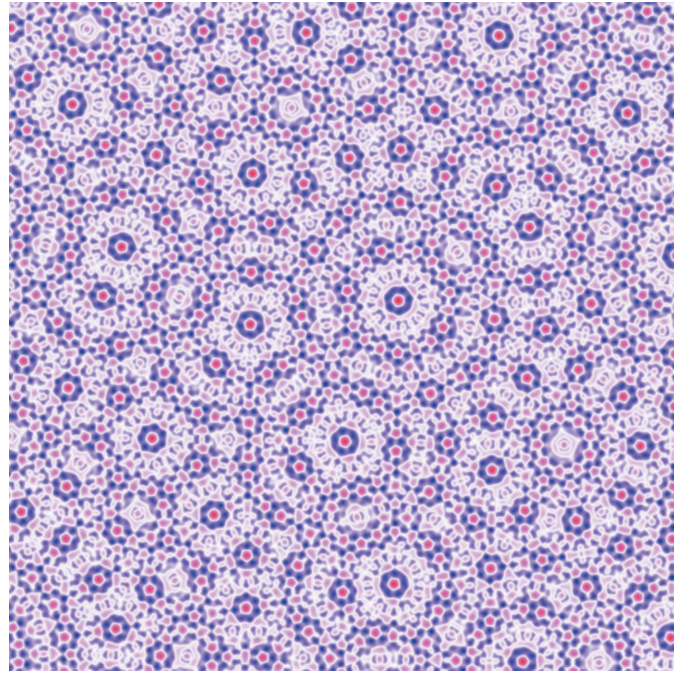

**Figure 13**
Predicted four-scale octadecagonal quasicrystal: blue and red shades correspond to negative field values $\phi \gtrsim -0.1615$ and positive values $\phi \lesssim 0.6008$, respectively. This quasicrystal has $\epsilon = -0.06$. At this $\epsilon$, the minimization of the quartic energy in equation (35$b$) gives $\tilde{\rho}_1 \simeq 0.02960$, $\tilde{\rho}_{q_1} \simeq 0.01234$, $\tilde{\rho}_{q_2} \simeq 0.01243$ and $\tilde{\rho}_{q_3} \simeq 0.01246$.

Rather than evaluating the integral (12) – used to calculate the contribution of the local term to the free energy – over space, this term can be summed differently by integrating over the set of possible values $\phi$ that $\rho(\mathbf{r})$ may take on,

$$\mathcal{F} = \int f(\phi) P(\phi) \, \mathrm{d}\phi, \qquad (36)$$

where

$$P(\phi) \propto \int \delta(\rho(\mathbf{r}) - \phi) \, \mathrm{d}\mathbf{r}, \qquad (37)$$

is normalized such that

$$\int P(\phi) \, \mathrm{d}\phi = 1, \qquad (38)$$

and $\delta$ is the Dirac delta function. $P(\phi)$ is the probability density function of $\rho(r)$ being $\phi$. Intuitively, it is essentially a histogram of the position-space $\rho(\mathbf{r})$ values obtained when $\mathbf{r}$ is selected by blindly throwing a dart at the entire Cartesian plane on which the structure is defined. This reformulation of the free energy is conceptually reminiscent of Lebesgue integration. It can also be thought of as taking a uniform-weight inner product of $f$ and $P$ over the vector space of real functions.

An intriguing analogy exists between the density distribution and the density of energy eigenstates of a quantum Hamiltonian for a single particle in a periodic potential. There, it is the energy dispersion, or band structure, $E(\mathbf{k})$, that plays the role of our density field $\rho(\mathbf{r})$. When the Hamiltonian is that of a nearest-neighbor tight-binding model for a particle hopping on a lattice, corresponding to one of our candidate structures, the band structure $E(\mathbf{k})$ assumes the same form taken by our density field $\rho(\mathbf{r})$ and the analogy becomes exact.[10] Consequently, the formal expression for the calculation of the density distribution in one dimension is similar to that of the density of states[11]

$$P(\phi) = \sum_{\rho(\mathbf{r}) = \phi} \frac{1}{|\nabla \rho(\mathbf{r})|}, \qquad (39)$$

where in $d$ dimensions the sum is replaced by an integral over the $d - 1$-dimensional equal-$\phi$ surface.

We can take the analogy one step further if we notice – using standard complex analysis – that the density distribution $P(\phi) = -\mathrm{Im}\{G(\phi)\}/\pi$, where

$$G(\phi) = \int \frac{1}{\phi - \rho(\mathbf{r}) + i0^+} \, \mathrm{d}\mathbf{r}. \qquad (40)$$

It turns out that this function $G(\phi)$ is the on-site lattice Green's function, obtained for the nearest-neighbor tight-binding Hamiltonian of a particle hopping on the corresponding lattice, with $\rho(\mathbf{r})$ replaced by $E(\mathbf{k})$. The real and imaginary parts of the lattice Green's function are related to each other by the Kramers–Kronig relations and so encode equivalent information about the crystal structure. The real part can be useful for evaluating the density distribution-based free-energy equation (36) for certain local free-energy density functions $f(\phi)$, such as the one in Section 7.

Many advanced mathematical approaches have been developed for evaluating these lattice Green's functions, including contour integrals (Ray, 2014), hypergeometric functions and Calabi–Yau differential equations (Guttmann, 2010), holonomic functions (Koutschan, 2013; Zenine *et al.*, 2015; Hassani *et al.*, 2016), and Chebyschev polynomials (Loh, 2017).

### 6.2. Rescaling and skewness of the density distribution

As for any normalized probability distribution (38), rescaling the field strength $\rho(\mathbf{r})$, and with it the width of the density distribution, merely rescales the distribution itself by the reciprocal factor, namely $P_\alpha(\alpha\phi) = P(\phi)/|\alpha|$. In the case of single-scale structures, whose overall field strength is determined by a single Fourier amplitude $\tilde{\rho}$, it is therefore sufficient to calculate the density distribution once for $P_{\tilde{\rho}=1}(\phi)$, and rescale later if necessary.

For two-scale structures, characterized by two Fourier amplitudes $\tilde{\rho}_1$ and $\tilde{\rho}_q$, a rescaling of $\rho(\mathbf{r})$ affects both amplitudes together, giving $P_{\tilde{\rho}_1, \tilde{\rho}_q}(\phi) = |\alpha| P_{\alpha\tilde{\rho}_1, \alpha\tilde{\rho}_q}(\alpha\phi)$. Thus, the density distributions differ for fields with different ratios $\tilde{\rho}_q/\tilde{\rho}_1$ of the amplitudes, but are otherwise independent of the overall scale of the field.

For all the structures relevant to us, $P(\phi)$ has compact support $[\phi_{\min}, \phi_{\max}]$ between the extreme values of $\rho(\mathbf{r})$, which are both finite, because the fields are all finite sums of harmonic functions. The value $\gamma = -\phi_{\max}/\phi_{\min}$ is a measure of the 'skewness' or 'lopsidedness' of the density distribution. It characterizes the imbalance between the ground state and the highest excited state of the corresponding tight-binding model. It is a useful measure that will serve us in what follows.

Because of the freedom to rescale the density distribution, for single-scale structures $\gamma$ can take on at most only two distinct values: $\gamma$ for positive $\tilde{\rho}$ and its inverse $1/\gamma$ for negative $\tilde{\rho}$. We need only consider positive $\tilde{\rho}$. On the other hand, with this assumption, $\gamma$ for two-scale structures varies continuously as a function $\gamma(\tilde{\rho}_q/\tilde{\rho}_1)$ of the ratio of the two Fourier amplitudes.

---

[10] This analogy is reminiscent of that between the shape of micelles in real space and electronic Fermi surfaces in momentum space, suggested by Lee *et al.* (2014) and discussed by Lifshitz (2014).

[11] See, for example, equation (8.63) of Ashcroft & Mermin (1976).

### 6.3. Numerical sampling of the field

One may need to resort to numerical sampling of the field $\rho(\mathbf{r})$ in order to generate the density distribution $P(\phi)$ when analytical methods for calculating equations (37) or (39) prove difficult. For periodic crystals this is readily achieved by uniformly sampling the unit cell of the crystal in both spatial directions. Quasicrystals lack periodicity, so this approach would, in principle, require a uniform sampling of the entire two-dimensional plane.

An alternative approach for the periodic case, which is easier to generalize to quasicrystals, is to remain at the origin of the two-dimensional plane and shift the field itself, until a full unit cell is sampled. This procedure samples the origin of the degenerate minimum free-energy states, which in the periodic case merely differ by a rigid translation within the unit cell.

One can sample the minimum free-energy states in terms of the Fourier coefficients of the fields by ensuring that the value of the free energy – like the one in equation (13) – does not change. This implies that one may generally shift the phases of the (complex) Fourier coefficients as long as the sum of these phases is zero for all possible structure invariants. This amounts to performing a Rokhsar–Wright–Mermin gauge transformation (Rokhsar *et al.*, 1988), as explained elsewhere (Lifshitz, 2011). Thus, one may freely shift the phases of the Fourier coefficients on wavevectors that are linearly independent over the integers. All the phase shifts of the remaining Fourier coefficients are then determined by the structure invariants. For periodic crystals in two dimensions there are two such independent phases. For the decagonal and dodecagonal quasicrystals of interest there are four independent phases. Shifting these phases uniformly from 0 to $2\pi$ yields the uniform sampling that we seek.[12]

### 6.4. Density distributions for the candidate phases

#### 6.4.1. Single-scale structures.
Trivially, $P_{\text{UNIF}}(\phi) = \delta(\phi)$.

We therefore begin with the single-scale lamellar field which, after scaling $\tilde{\rho}$ to unity, is given by $\rho(\mathbf{r}) = 2\cos x$. Because $|\phi|$ never exceeds two, $P_{\text{LAM}}(\phi)$ vanishes when $|\phi| > 2$. We express it analytically between these bounds using equation (39), by substituting $-2\sin(x)\hat{x}$ for $\nabla\rho(\mathbf{r})$ and $\cos^{-1}(\phi/2)$ for $x$ and normalizing according to equation (38). This gives

$$P_{\text{LAM}}(\phi) = \begin{cases} 0 & \phi \leq -2, \\ \dfrac{1}{\pi(4-\phi^2)^{1/2}} & -2 \leq \phi \leq 2, \\ 0 & 2 \leq \phi. \end{cases} \quad (41a)$$

This calculation is demonstrated schematically in Fig. 14.

While calculating $P_{\text{LAM}}$ is straightforward, doing so for $P_{\text{HEX}}$ is quite difficult. The mathematics necessary to do so was worked out by Ramanujan (1914) using one of his theories of elliptic functions to alternative bases. The connection to lattice Green's functions was introduced by Horiguchi (1972). We simply provide the result,

$$P_{\text{HEX}}(\phi) = \begin{cases} 0 & \phi < -3, \\[2mm] \dfrac{2K\left(\dfrac{16(\phi+3)^{1/2}}{8(\phi+3)^{1/2}-\phi^2+12}\right)}{\pi^2[8(\phi+3)^{1/2}-\phi^2+12]^{1/2}} & -3 \leq \phi < -2, \\[4mm] \dfrac{2K\left(1-\dfrac{16(\phi+3)^{1/2}}{8(\phi+3)^{1/2}+\phi^2-12}\right)}{\pi^2[8(\phi+3)^{1/2}+\phi^2-12]^{1/2}} & -2 < \phi \leq 6, \\[4mm] 0 & 6 < \phi, \end{cases} \quad (41b)$$

where $K$ is the complete elliptic integral of the first kind.

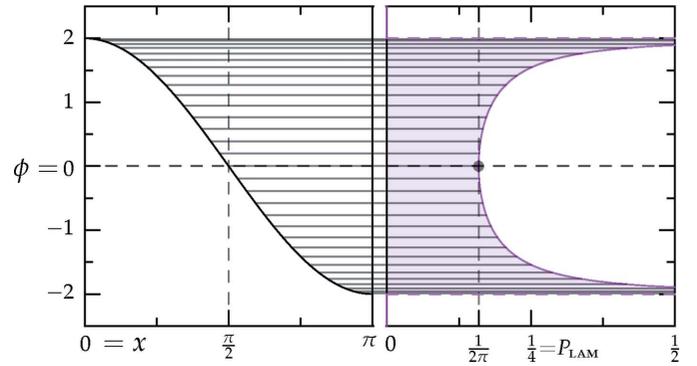

**Figure 14**
Graphical evaluation of the lamellar density distribution: the left-hand panel shows a half-period of the sinusoidal lamellar spatial structure. The horizontal lines coming off it are evenly spaced in the horizontal direction. Their vertical density determines the distribution in the right-hand panel. Note how the stationary regions of the structure, where the gradient vanishes at 0 and $\pi$, lead to the inverse square root Van Hove singularities in the density distribution at $|\phi| = 2$.

These density distributions are plotted in Fig. 15, showing that the lamellar phase is unskewed with $\gamma_{\text{LAM}} = 1$ as expected, while the hexagonal phase is skewed, with $\gamma_{\text{HEX}} = 2$.

#### 6.4.2. Two-scale structures.
The density distributions of the decagonal and dodecagonal fields are calculated numerically by measuring them at the origin, as explained earlier, while sampling the set of all degenerate minimum free-energy states *via* appropriate phase shifts of the harmonic functions. The phase-shifted fields are given by

$$\begin{aligned} \rho_{\text{DEC}}(\mathbf{r} = 0; \{\chi_i\}) = \\ 2\tilde{\rho}_1[\cos\chi_1 + \cos\chi_2 + \cos\chi_3 + \cos\chi_4 \\ + \cos(\chi_1 + \chi_2 + \chi_3 + \chi_4)] \\ + 2\tilde{\rho}_q[\cos(\chi_1 + \chi_2) + \cos(\chi_2 + \chi_3) + \cos(\chi_3 + \chi_4) \\ + \cos(\chi_1 + \chi_2 + \chi_3) + \cos(\chi_2 + \chi_3 + \chi_4)], \end{aligned} \quad (42a)$$

and

---

[12] Equivalently, one can think of this process as uniformly sampling a unit cell of the higher-dimensional periodic structure through which the quasicrystal can be constructed as a slice at an irrational slope (Senechal, 1995).

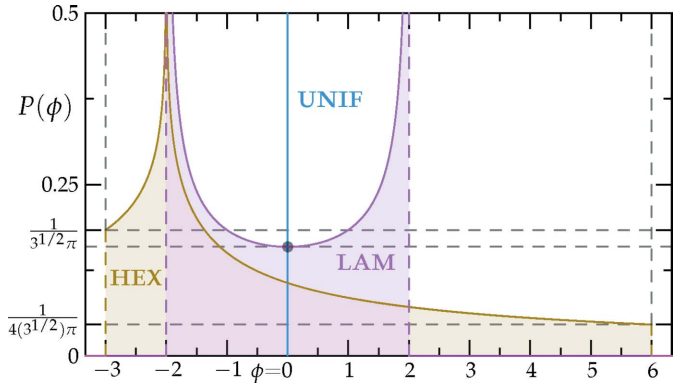

**Figure 15**
Density distributions of the uniform, lamellar and hexagonal structures: note the Van Hove singularities (i) the inverse root singularities of the lamellar distribution at $\phi_{\min} = -2$ and $\phi_{\max} = 2$, (ii) the logarithmic singularity of the hexagonal distribution at $\phi = -2$, and (iii) the discontinuous jumps from $1/(3^{1/2}\pi)$ and $1/[4(3^{1/2})\pi]$ to zero at $\phi_{\min} = -3$ and $\phi_{\max} = 6$, respectively. The skewness of the latter two distributions is given by $\gamma_{\text{LAM}} = 1$ and $\gamma_{\text{HEX}} = 2$.

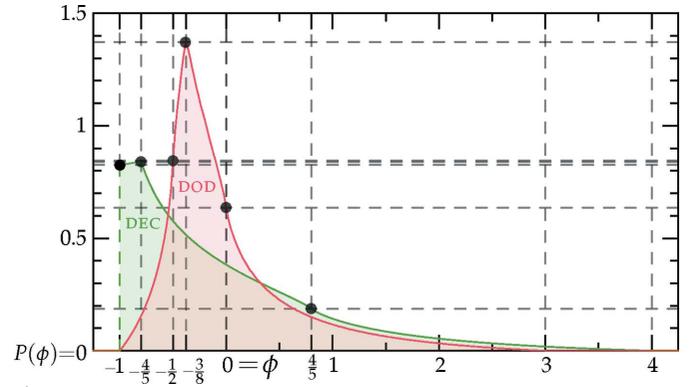

**Figure 16**
Density distributions for decagonal and dodecagonal quasicrystals: for these quasicrystals, $\gamma$ is maximized when $\tilde{\rho}_1 = \tilde{\rho}_q = 1/5$ or $1/8$, respectively. Observe that $\gamma_{\text{DEC}} = 4$ and $\gamma_{\text{DOD}} = 3$, and note the interior Van Hove singularities at $\pm 4/5$ and $-1/2$, $-3/8$ and $0$, and the zeroth-order discontinuity in the decagonal distribution at $\phi_{\min} = -1$. This final Van Hove singularity is analyzed in Section 6.5 and is a key factor leading to the decagonal structure's stability under the BDL model.

$$\rho_{\text{DOD}}(\mathbf{r} = 0; \{\chi_i\}) =$$
$$2\tilde{\rho}_1[\cos \chi_1 + \cos \chi_3 + \cos (\chi_1 + \chi_3)$$
$$+ \cos \chi_2 + \cos \chi_4 + \cos (\chi_2 + \chi_4)]$$
$$+ 2\tilde{\rho}_q[\cos (\chi_1 + \chi_2) + \cos (\chi_3 + \chi_4)$$
$$+ \cos (\chi_1 + \chi_2 + \chi_3 + \chi_4)$$
$$+ \cos (\chi_1 + \chi_2 + \chi_3) + \cos (\chi_4 - \chi_1)$$
$$+ \cos (\chi_2 + \chi_3 + \chi_4)]. \quad (42b)$$

We call $\phi_1$ the value of $\rho(\mathbf{r} = 0; \{\chi_i\})$ when $\tilde{\rho}_1 = 1$ and $\tilde{\rho}_q = 0$ and likewise call $\phi_q$ the value when $\tilde{\rho}_1 = 0$ and $\tilde{\rho}_q = 1$, so that $\phi = \tilde{\rho}_1\phi_1 + \tilde{\rho}_q\phi_q$. We numerically sample the ordered pairs $(\phi_1, \phi_q)$ uniformly over $\chi_i \in [0, 2\pi)$, with 96 sampling points along each phase, to give a total of roughly 85 million samples. If we write this distribution function as $P(\phi_1, \phi_q)$, then

$$P_{\tilde{\rho}_1, \tilde{\rho}_q}(\phi) = \iint \delta(\tilde{\rho}_1\phi_1 + \tilde{\rho}_q\phi_q - \phi) P(\phi_1, \phi_q) \, \mathrm{d}\phi_1 \, \mathrm{d}\phi_q. \quad (43)$$

Upon numerically maximizing the skewness parameter $\gamma$ over the ratio of amplitudes $\tilde{\rho}_q/\tilde{\rho}_1$ for the decagonal and dodecagonal phases, we find that the optimal ratio is unity, where $\tilde{\rho}_1 = \tilde{\rho}_q > 0$, in both cases. The density distributions obtained for this ratio are shown in Fig. 16, where it can be seen that the decagonal and dodecagonal phases have maximal $\gamma$ values of 4 and 3, respectively.

The density distributions in Figs. 15 and 16, along with equation (36), allow us to understand the stability of the candidate phases more generally than before. The uniform phase simply has $\mathcal{F}_{\text{UNIF}} = f(0)$. The lamellar phase can potentially dominate this by heavily sampling the local free-energy density $f(\phi)$ far from $\phi = 0$. For example, a local free-energy density with a strongly negative quadratic component would likely favor the lamellar phase. Indeed, this is what we observe in Section 3.2.1 when $\epsilon$ exceeds $(6^{3/2} + 14)/15$ and the free energy of the lamellar phase is lower than that of the hexagonal phase. On the other hand, the lopsided nature of the hexagonal structure allows it to take advantage of the odd components of the free-energy density.

Similarly, quasicrystalline structures also attain stability through the skewness of their density distributions. In particular, their extremes can be more lopsided than those of the hexagonal phase. In other words, they have $\gamma > \gamma_{\text{HEX}} = 2$. In Section 8, we use this feature of quasicrystals to stabilize arbitrarily high-order structures using only two length scales.

### 6.5. Van Hove singularities

As shown in Figs. 15 and 16, the density distributions exhibit a variety of Van Hove singularities (Van Hove, 1953). Note the inverse square root Van Hove singularities exhibited by the lamellar structure and the zeroth-order step discontinuities and logarithmic singularities in the hexagonal density histogram. The logarithmic singularity at $\phi = -2$ is due to the corresponding stationary point in the structure function being a saddle, to leading order, unlike the extrema which lead to the step discontinuities.

By numerically minimizing the magnitude of the gradients of the effective four-dimensional periodic functions [equations (42a) and (42b)],

$$\sum_{i=1}^{4} \left(\frac{\partial \rho}{\partial \chi_i}\right)^2, \quad (44)$$

we can identify Van Hove singularities at $\phi_{\min}$ and $\phi_{\max}$, at $\pm 4/5$ in the decagonal structure, and at $-1/2$, $-3/8$ and 0 in the dodecagonal structure.

With the exception of the discontinuity in the decagonal distribution at $\phi_{\min} = -1$, all of the quasicrystal Van Hove singularities are first order. This zeroth-order Van Hove singularity turns out to be crucial for stabilizing the decagonal structure under the BDL model in Section 7 and can be understood in terms of the spatial Hessian of the effective four-dimensional function. One of the phase-shifted structures that has this minimum of $-1$ at its origin is given by $\{\chi_i\}_{\min} = \{\sec^{-1}(-4), 0, 2\sec^{-1}(4), 0\}$. At this point, the Hessian is

$$\left.\frac{\partial^2 \rho_{\text{DEC}}}{\partial \chi_i \partial \chi_j}\right|_{\{\chi_i\}_{\min}} = \frac{1}{4}\begin{pmatrix} 8 & 6 & 4 & 2 \\ 6 & 12 & 18 & 9 \\ 4 & 18 & 32 & 16 \\ 2 & 9 & 16 & 8 \end{pmatrix}. \qquad (45)$$

The nullity, or the rank of the kernel, of this matrix is two. In this case, this indicates that the manifold of the minima is two-dimensional. This generates a quadratic minimum of effective dimension two, the rank of the Hessian. A quadratic $d$-dimensional minimum generates a singularity of order $d/2 - 1$, so the decagonal density distribution has a zeroth-order discontinuity at its minimum.

As explained by Van Hove (1953), topological considerations in Morse theory require the existence of a certain number of stationary points with each quadratic signature, although degenerate stationary points such as the one analyzed in the previous paragraph complicate the situation.

### 6.6. A lower bound on the LP free energy

Using the language of density distributions, we can calculate a lower bound for the free energy in the LP model. Clearly, if it were not for the infinite-$c$ penalty at $\mathbf{k} = 0$, which requires the average density to be zero, the best possible density distribution would have a single delta-function peak, corresponding to a uniform field $\rho(\mathbf{r}) = \phi_0$ that minimizes the local free-energy density (7) everywhere. To satisfy the zero-average requirement we must add a compensating delta function peak at some negative value $\phi_\ell < 0$, and possibly allow the positive value $\phi_r > 0$ to shift away from $\phi_0$. This yields a field with sharp boundaries between two allowed values and a density distribution of the form $P(\phi) = P_\ell \delta(\phi - \phi_\ell) + P_r \delta(\phi - \phi_r)$.

Given sufficient harmonics, structures of arbitrary symmetry can attain this sharp form. Of course, with too many allowed length scales, the physical requirements on the length-scale selectivity $c$ are stricter, and even so the set of competing candidate phases can increase, so the results below provide only an extreme theoretical lower bound on the free energy.

All that is left is to minimize the free energy (36) under the constraints of zero-averaging, $P_\ell \phi_\ell + P_r \phi_r = 0$, and the normalization $P_\ell + P_r = 1$ of the density distribution. Solving for the left-hand side variables gives $P_\ell = 1 - P_r$ and $\phi_\ell = (P_r \phi_r)/(P_r - 1)$. Substituting them into $P(\phi)$ and minimizing the free energy (36) over $P_r$ and $\phi_r$ gives

$$P_r = \frac{1}{2} - \frac{1}{2(9\epsilon + 3)^{1/2}}, \qquad (46)$$

and

$$\phi_r = \frac{(9\epsilon + 3)^{1/2} + 1}{3}. \qquad (47)$$

Essentially, we have fitted the right peak into the wells of the solid colored lines in Fig. 2, while remembering that it must be balanced out by a corresponding peak at negative $\phi$. This gives a lower bound on the free energy of $\mathcal{F} \geq -(9\epsilon + 2)^2/324$.

Note that this implies that $\epsilon$ must be greater than $-2/9$ to allow for the possibility of structures with negative free energy. This 'forbidden zone' is shown as the grayed-out regions in Figs. 4–6. Furthermore, the lowest possible metastability bound is $\epsilon = -1/3$.

## 7. Mean-field theory for soft interacting particles

### 7.1. The Barkan–Diamant–Lifshitz model

Equipped with the density distribution method and the ability to calculate free energies with non-polynomial local terms, we can perform a more detailed and informed analysis of the coarse-grained free energy (3) used in the BDL model (Barkan *et al.*, 2011), which contains a local logarithmic entropy term. Assuming a sufficiently dense system of soft particles that interact *via* a Fourier transformable isotropic pair potential $\mathcal{U}(r)$ – implying that it does not diverge at a higher order than the usual $1/r$ electrostatic potential as the particles get closer together – one can express the BDL coarse-grained free energy in the form of equation (9) with

$$\tilde{V}(k) = \frac{\tilde{\mathcal{U}}(k) - \tilde{\mathcal{U}}_{\min}}{8\pi^2 |\tilde{\mathcal{U}}_{\min}|}, \qquad (48)$$

and

$$f_{\text{CG}}(\phi) = T\left[(\phi + 1)\ln(\phi + 1) - \phi\right] - \frac{\phi^2}{2}, \qquad (49)$$

where $\tilde{\mathcal{U}}(k)$ is the Hankel transform of $\mathcal{U}(r)$ and $\tilde{\mathcal{U}}_{\min}$ is its minimum value. The temperature $T$ is measured here in units of the spinodal temperature $T_{\text{sp}} = -\bar{c}\tilde{\mathcal{U}}_{\min}/k_B$, where $\bar{c}$ is the average number density of the particles and $k_B$ is the Boltzmann constant. Recall that $T_{\text{sp}}$ is the lower metastability boundary of the uniform liquid phase, below which the system must become ordered, and note that the minimum $\tilde{\mathcal{U}}_{\min}$ of $\tilde{\mathcal{U}}(k)$ must be negative for this temperature to be positive.[13] Finally, the value $\phi$ of the field $\rho(\mathbf{r})$ is here constrained to $\phi > -1$ by the fact that the number density $c(\mathbf{r}) = \bar{c}\left[\rho(\mathbf{r}) + 1\right]$ of the particles cannot be negative. This 'vacuum constraint' ensures that the logarithm in equation (49) does not diverge.

BDL proceeded to take the fourth-order Taylor expansion of $f_{\text{CG}}$ to obtain

$$f_4(\phi) = \frac{T - 1}{2}\phi^2 - \frac{T}{6}\phi^3 + \frac{T}{12}\phi^4, \qquad (50)$$

and mapped this resulting quartic free energy onto the LP free energy, giving them a rough estimate of the physical parameters that might stabilize the different targeted structures, based on the LP results. By rescaling $\phi$ and $\mathcal{F}$, we can make $f_4$ equivalent to the LP form in equation (7). The correspondence between $T$ and $\epsilon$ necessary to do so is then given by $T = 4/(3\epsilon + 4)$. Note that the range $-4/3 < \epsilon < 0$ corresponds to scaled temperature values of $T > 1$ above the spinodal decomposition, and that positive values of $\epsilon$ correspond to values of $T < 1$ below it. The cases of $\epsilon \leq -4/3$ and $T < 0$ are unphysical for the coarse-grained free-energy model.

---

[13] It must also be noted that BDL denote the spinodal temperature as $T_c$, while Barkan *et al.* (2014), who confirmed the BDL results using molecular dynamics simulations, denote it as $T_{\text{sp}}$ but leave the temperature $T$ unscaled.

The success of the estimates obtained by BDL through this mapping were somewhat fortuitous, as a comparison between even properly rescaled plots of $f_{LP}$ and $f_{CG}$ in Fig. 2 reveals that they are very different outside of the radius of convergence $|\phi| > 1$. As the transition from the uniform liquid to the ordered state is first order, the field $\rho(\mathbf{r})$ generally contains regions with large values, making $f_4$ a poor approximation even at the transition. It is therefore important that we can now evaluate the exact local free energy $f_{CG}$. As for the evaluation of the nonlocal term of the free energy, in order to make the analytical calculations more tractable, we again work in the limit of exact length-scale selection – analogous to the infinite-$c$ limit of the LP model – corresponding here to the small $\tilde{\mathcal{U}}_{min}$ and therefore small $T_{sp}$ limit. As seen below, we indeed find important differences between the behaviors of the LP and BDL models.

## 7.2. Single-scale structures

For the single-scale lamellar and hexagonal phases, substituting the density distributions of equations (41a) and (41b) and the logarithmic local free-energy of equation (49) into the free-energy equation (36) gives

$$\mathcal{F}_{LAM}(\tilde{\rho}; T) = -\tilde{\rho}^2 + T\left\{\ln\left[(1 - 4\tilde{\rho}^2)^{1/2} + 1\right] - (1 - 4\tilde{\rho}^2)^{1/2} - \ln 2 + 1\right\},$$

(51a)

and

$$\mathcal{F}_{HEX}(\tilde{\rho}; T) = -3\tilde{\rho}^2 + T(6\tilde{\rho} + 1)^2$$
$$\times \int_0^1 \frac{\left[{}_2F_1\left(\frac{1}{3}, \frac{2}{3}; 1; \frac{27\tilde{\rho}^2 x^2(8\tilde{\rho}x - 6\tilde{\rho} - 1)}{(6\tilde{\rho}x - 6\tilde{\rho} - 1)^3}\right) - 1\right](x - 1)}{(6\tilde{\rho}x - 6\tilde{\rho} - 1)x^2}\,dx,$$

(51b)

where ${}_2F_1$ is a hypergeometric function.[14]

Now, minimizing $\mathcal{F}_{LAM}$ over $\tilde{\rho}$ yields

$$\mathcal{F}_{LAM}(T) = \begin{cases} T - T\ln(2) - \frac{1}{4} & 0 \leq T \leq \frac{1}{2}, \\ T[\ln(T) - T + 1] & \frac{1}{2} \leq T \leq 1, \\ 0 & 1 \leq T, \end{cases}$$

(52)

which is shown in Fig. 17. At absolute zero, the free energy is exactly $-1/4$. In the first temperature range, where $\mathcal{F}_{LAM}$ is linear in $T$, $\tilde{\rho} = 1/2$, which is the maximum value it can obtain without violating the vacuum constraint. Increasing it further would violate the non-negative density condition. At the transition to the second temperature range, the free energy is $(1 - 2\ln 2)/4$. From here, as $T$ increases to unity, $\tilde{\rho}$ decreases

[14] Alternatively, one could evaluate these expressions using a double integral of an expression involving the analytically continued real part of the lattice Green's function (40) and no logarithms, instead of a single integral with a logarithm.

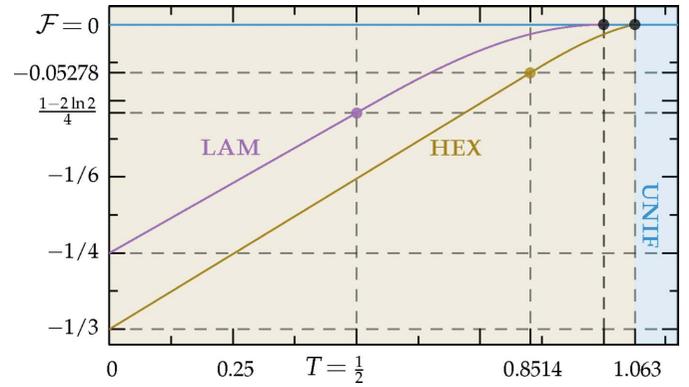

**Figure 17**
Free energies of the single-scale periodic structures in the BDL model: in this model, the free energy of the hexagonal structure is always lower than that of the lamellar structure, so we need not consider either. The hexagonal structure is the equilibrium phase up to $T \simeq 1.063$, where it undergoes a first-order transition to the uniform liquid phase. Colored dots correspond to temperatures below which the Fourier amplitude $\tilde{\rho}$ reaches its maximum value and the free energy becomes a linear function of the temperature.

to zero. At $T = 1$, a second-order phase transition to the uniform phase occurs.

The result of minimizing $\mathcal{F}_{HEX}$, which is obtained numerically, is also shown in Fig. 17, displaying similar behavior. At absolute zero, the free energy is exactly $-1/3$. For $0 \leq T \lesssim 0.8514$, $\tilde{\rho}$ has its maximum allowed value, which is $1/3$, and the free energy is linear in $T$. When $T \simeq 0.8514$, $\mathcal{F} \simeq -0.05278$, and in the next temperature range $\tilde{\rho}$ decreases monotonically to $\sim 0.1923$ at $T \simeq 1.063$, at which point the system undergoes a first-order transition to the uniform phase.

Note that the hexagonal phase always has a lower free energy than the lamellar phase. Therefore, we do not need to consider the single-scale lamellar candidate when evaluating the stability of quasicrystals in the BDL model, as it is never the equilibrium phase. This is qualitatively different from the behavior with the quartic local free energy $f_4$, analogous to that of the LP model, where the lamellar phase would be expected to take over at $T \leq (34 - 6^{3/2})/47 \simeq 0.4107$.

## 7.3. Two-scale structures

In the original LP model, one can set $q$ to $k_4$, $k_6$ and $k_\infty$ to stabilize two-scale square, hexagonal and lamellar superstructures, respectively. Unsurprisingly, the BDL model can also do this, as demonstrated in Fig. 18, and it stabilizes these two-scale periodic phases all the way down to absolute zero. Their free energies are calculated using the same techniques as the quasicrystalline structures treated below, but we omit a detailed analysis of their behavior.

Again, as in the LP model, setting $q$ to $k_8$ and $k_{10}$ fails to stabilize octagonal and decagonal quasicrystals, as their free energies, shown in Fig. 18, are always greater than that of the single-scale hexagonal phase. Only decagonal and dodecagonal structures with $q = k_5$ and $k_{12}$ occur as stable quasicrystalline states in the BDL model. Their free energies are calculated using the sampled distribution of ordered pairs

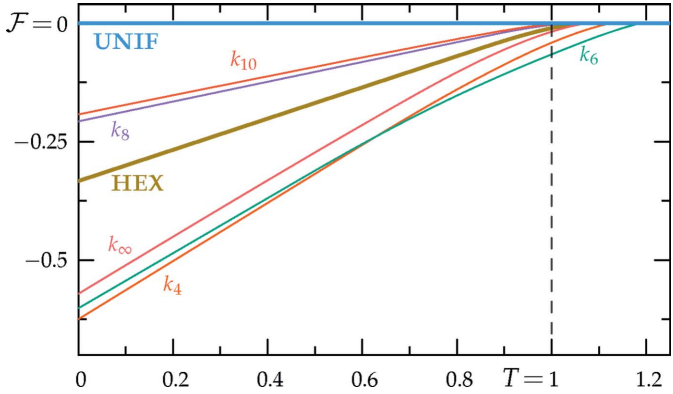

**Figure 18**
Free energies of the two-scale periodic and unstable quasiperiodic structures in the BDL model: plotted for comparison are the free energies of the uniform and single-scale hexagonal phases. $q$ takes the values $k_4$, $k_6$, $k_8$, $k_{10}$ and $k_\infty$, as labeled. As in the LP model, the free energies of the two-scale square ($k_4$), hexagonal ($k_6$), lamellar ($q = k_\infty$), decagonal ($k_{12}$) and dodecagonal ($k_{12}$) structures are lower than that of single-scale hexagonal structure, but unlike the LP model the decagonal structure remains the equilibrium phase down to zero temperature, without undergoing a second phase transition. On the other hand, still in line with the LP model, the free energies of the octagonal ($k_8$) and decagonal ($k_{10}$) quasicrystals are always higher than that of the single-scale hexagonal structure, and are therefore never the equilibrium phase.

$P(\phi_1, \phi_q)$ as explained in Section 6.4.2, and plotted in Fig. 19 as functions of $T$.

We would like to emphasize that the minimization of the free energy with respect to $\tilde{\rho}_1$ and $\tilde{\rho}_q$ is performed subject to the vacuum constraint, which is more difficult to take into account for the two-scale structures. Each $(\phi_1, \phi_q)$ pair from Section 6.4.2 not only gives a small free-energy contribution, but also imposes the linear constraint $\phi_1\tilde{\rho}_1 + \phi_q\tilde{\rho}_q \geq -1$ on the values of $\tilde{\rho}_1$ and $\tilde{\rho}_q$. The problem of finding the intersection of these half-planes is dual to that of finding the convex hull of the $(\phi_1, \phi_q)$ points (de Berg *et al.*, 2008). If $(\phi_{1A}, \phi_{qA})$ and $(\phi_{1B}, \phi_{qB})$ are adjacent extremal points on the convex hull, then the point

$$\frac{(\phi_{qA} - \phi_{qB}, \phi_{1B} - \phi_{1A})}{\phi_{1A}\phi_{qB} - \phi_{1B}\phi_{qA}},\tag{53}$$

is a vertex of the polygonal boundary of the set of allowed $(\tilde{\rho}_1, \tilde{\rho}_q)$ values that do not violate the vacuum constraint. The feasible sets for the decagonal and dodecagonal quasicrystals are displayed in Fig. 20. As shown there, we are able to find exact values for all the polygonal vertices bounding the regions. A simple constrained descent algorithm is used to minimize the free energy over these convex sets.

For the decagonal phase with $q = k_5$, a portion of which is shown in Fig. 21, as the temperature is lowered the system undergoes a first-order phase transition from the uniform liquid to the quasicrystal at $T \simeq 1.125$, at which point $\tilde{\rho}_1 = \tilde{\rho}_q = 0.1352$. These $\tilde{\rho}$'s increase together until $T \simeq 0.8977$, where they reach their maximal allowed value of $1/5$. At this point, the free energy of the decagonal phase is approximately $-0.06746$. This continues to be the equilibrium phase all the way down to absolute zero, where the free energy becomes exactly $-2/5$.

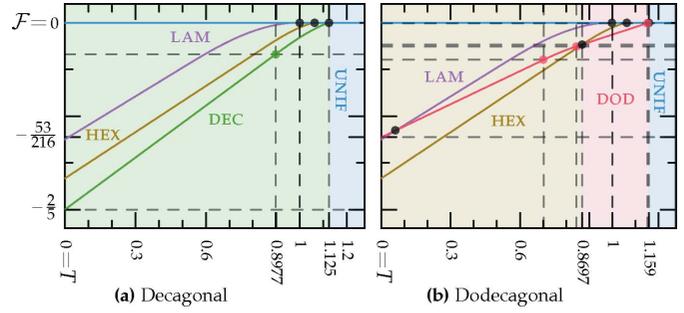

**Figure 19**
Free energies of the stable two-scale quasiperiodic structures in the BDL model: the decagonal structure ($k_5$) is the equilibrium phase for $T \lesssim 1.125$ and the dodecagonal structure ($k_{12}$) is the equilibrium phase for $0.8697 \lesssim T \lesssim 1.159$. Black dots mark the phase transitions, while colored dots mark transitions between linear and nonlinear regimes of the free energy, where the Fourier amplitudes $\tilde{\rho}_1$ and $\tilde{\rho}_q$ reach a stationary pair of values.

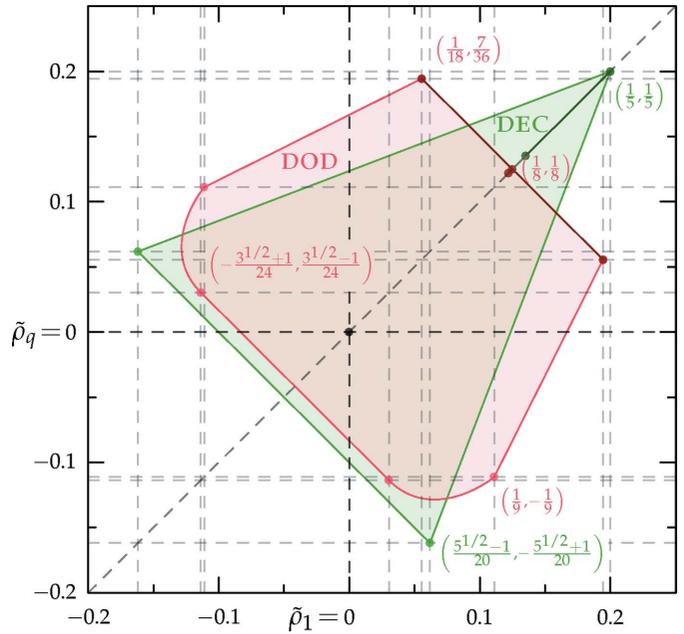

**Figure 20**
Feasible $(\tilde{\rho}_1, \tilde{\rho}_q)$ sets for decagonal and dodecagonal quasicrystals in the BDL model: these are the convex regions over which the Fourier coefficients can vary without violating the vacuum constraint. There is a mirror line given by $\tilde{\rho}_1 = \tilde{\rho}_q$. Solid dark lines in the first quadrant show the path of the minimizing values of $(\tilde{\rho}_1, \tilde{\rho}_q)$ as the temperature is changed. While the polygonal vertices are exact, we believe that there is no simple expression for the two smooth curves in the dodecagonal feasibility boundary. Note that the peakedness of the decagonal structure jutting far into the first quadrant essentially explains its surprising stability in the BDL model. Similar shapes exist for the additional phases considered in Fig. 18, but are omitted here.

For the dodecagonal phase with $q = k_{12}$, the first-order transition from the uniform liquid occurs at $T \simeq 1.159$, where $\tilde{\rho}_1 = \tilde{\rho}_q \simeq 0.1220$. At $T \simeq 1.152$, the free energy $\mathcal{F} \simeq -1.123 \times$ $10^{-3}$ and the $\tilde{\rho}$'s reach their maximum allowed value of $1/8$, and the free energy as a function of temperature enters a linear regime. The $\tilde{\rho}$'s remain at $1/8$ until $T \simeq 0.8697$, at which point the hexagonal phase takes over at a free energy of approximately $-0.04697$. Regardless, if we continue to restrict our attention to the dodecagonal structure, it undergoes a second-order phase transition-like event where the $\mathbb{Z}_2$ symmetry $\tilde{\rho}_1 \leftrightarrow \tilde{\rho}_q$ is broken at $T \simeq 0.8446$ and $\mathcal{F} \simeq -0.05086$. After this point, the $\mathcal{F}$'s move along their maximum allowed sum line $\tilde{\rho}_1 + \tilde{\rho}_q = 1/4$ until $T \simeq 0.7022$, where they land in either of the two degenerate states $(\tilde{\rho}_1, \tilde{\rho}_q) = (7/36, 1/18)$ or $(1/18, 7/36)$ with a free energy of approximately $-0.07973$. The structure remains in one of these two minima until absolute zero, where the free energy becomes exactly $-53/216$.

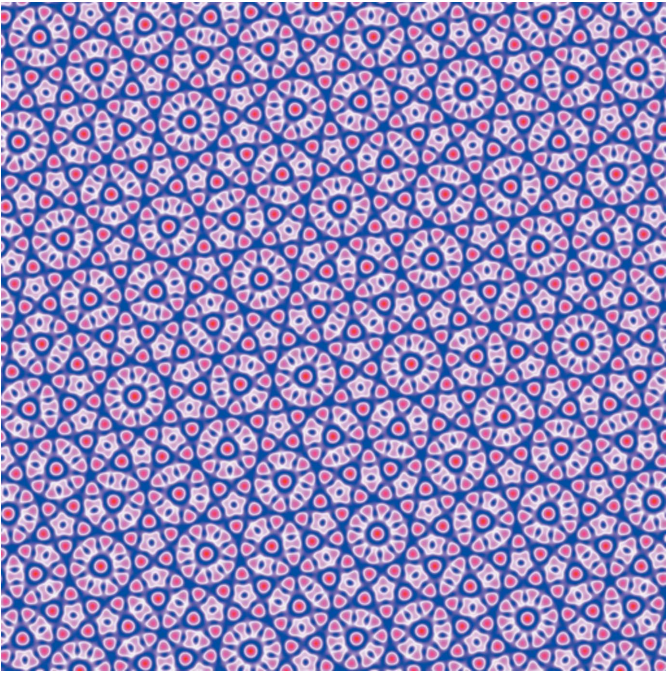

**Figure 21**
Predicted decagonal quasicrystal with $\tilde{\rho}_1 = \tilde{\rho}_q = 1/5$: red shades correspond to positive field values $\phi \leq 4$, whereas blue shades represent negative values which just barely scrape against the vacuum when $\phi = -1$. Note the abundance of blue or white areas which are interspersed with bright red spots. This provides the skewness which makes this structure so stable in the BDL model.

### 7.4. Skewness and the vacuum constraint

While the dodecagonal quasicrystal shows qualitatively similar stability under the LP and BDL models, the decagonal quasicrystal exhibits significantly different behavior, showing surprisingly robust stability in the BDL model. Until now, it was understood that soft quasicrystals are stabilized by three-body (or more generally odd-body) interactions that break the $\mathbb{Z}_2$ symmetry of the free energy $\mathcal{F}$, namely, the $\phi \rightarrow -\phi$ symmetry of $f(\phi)$. However, the fact that the decagonal quasicrystal dominates even at $T = 0$, where there are no three-body interactions, demonstrates that this cannot be the whole story.

The primary quasicrystal stabilizing factor in the BDL model is the important $\phi \geq -1$ vacuum constraint, which is a very effective alternative way of breaking the $\mathbb{Z}_2$ symmetry of $\mathcal{F}$. The decagonal structure with $\tilde{\rho}_1 = \tilde{\rho}_q$ possesses the most lopsided of any of the density distributions examined, giving it a $\gamma$ skewness of four, and allows the decagonal phase to take maximal advantage of the the highly negative $f_{CG}(\phi)$ values at large $\phi$'s without violating the vacuum constraint. The important Van Hove singularity at the vacuum minimum, which allows this high skewness to occur, is analyzed in Section 6.5. Note that the structure itself, shown in Fig. 21, contains well separated very highly peaked positive red spots in a shallow blue sea of negative values. This argument also explains the result of the molecular dynamics simulations performed by Barkan *et al.* (2014) with $q = k_s$, showing that the decagonal quasicrystal remains stable as $T$ is lowered to absolute zero, without undergoing a transition to the hexagonal phase as would be predicted by a naïve correspondence to the LP model.

## 8. Higher-order quasicrystals

### 8.1. Skewness of the 6*n*-fold two-scale structures

Finally, we examine the $\gamma$ skewness of the density distributions of quasicrystals of order $6n$ for arbitrary $n \geq 2$. Then, using the information obtained, we judiciously design an artificial local free-energy function $f(\phi)$ that allows for the stabilization of quasicrystals of these orders. While the local free energies previously used by Lifshitz & Petrich (1997) and Barkan *et al.* (2011) were physically justified by entropic considerations and their truncated polynomial expansions, the one we construct below is engineered with the sole goal of stabilizing higher-order quasicrystal phases. However, as we argue below, it might not be impossible to design a physical system with sufficiently similar behavior to stabilize some of these higher-order quasicrystals, particularly when $n$ is not too large.

We first demonstrate that, for all $n$, two-scale $6n$-fold crystals, like the ones shown in Figs. 3(*j*) and 3(*m*) for $n = 1$ and 2, all have their skewness $\gamma$ from Section 6.2 greater than two, when the two amplitudes $\tilde{\rho}_1$ and $\tilde{\rho}_q$ are equal. We restrict our attention to this case and scale both $\tilde{\rho}_1$ and $\tilde{\rho}_q$ to unity.

By inverse Fourier transforming its momentum-space representation according to equation (8), the position-space field representing this quasicrystal can be written as

$$\rho(\mathbf{r}) = \sum_{j=1}^{6n} \exp\left[i(\mathbf{k}_j \cdot \mathbf{r} + \chi_j)\right] + \exp\left\{i\left[(\mathbf{k}_j + \mathbf{k}_{j+1}) \cdot \mathbf{r} + \chi_{j,j+1}\right]\right\},$$

(54)

where the wavevectors

$$\mathbf{k}_j = \cos\left(\frac{j}{6n}\right)\hat{x} + \sin\left(\frac{j}{6n}\right)\hat{y}, \qquad (55)$$

have length $|\mathbf{k}_j| = 1$, and the sum of two consecutive vectors has a length $|\mathbf{k}_j + \mathbf{k}_{j+1}| = q = k_{6n}$.[15]

In principle, the additional phases $\chi_j$ and $\chi_{j,j+1}$ are free to vary so as to minimize the free energy (12), yet for similar arguments mentioned in Section 3.1.1 there is always a representative structure, within the set of all degenerate minimum free-energy states, for which the phases within each circle are all equal and can be taken to be 0 or $\pi$. We limit our attention to structures where the phases are the same on both circles, taking them all to be zero without any further loss of generality, owing to our freedom to change the sign of the cubic term in $f(\phi)$ accordingly. With this choice, with all the $\chi$'s set to zero, the field obtains its maximum value $\phi_{\max} = 12n$ at the origin. We now show that $\phi_{\min} > -6n$ so that $\gamma > 2$.

As explained in Section 6.3, we sample the field by staying at the origin $\mathbf{r} = 0$ and shifting the $\chi$ phases, while keeping the structure invariants constant, thereby performing Rokhsar–Wright–Mermin (Rokhsar *et al.*, 1988) gauge transformations. This requires the sum of the phase shifts at wavevectors that add to zero to vanish, and immediately implies that $\chi_{j,j+1} \equiv \chi_j + \chi_{j+1}$, where '$\equiv$' stands for equality modulo $2\pi$. The phase shifts on the outer circle are all fixed by the choice of shifts on the inner circle.

In addition, within the inner circle, each vector and its negative impose the constraint $\chi_j + \chi_{j+3n} \equiv 0$, and each triplet of wavevectors adding to zero imposes the constraint $\chi_j + \chi_{j+2n} \equiv \chi_{j+n}$. This leaves at most two independent phases on each of the $n$ sextets forming the inner circle that still need to satisfy additional constraints imposed by each additional prime divisor of $n$ other than two or three. The resulting number of independent phase shifts is given by $\Phi(6n)$, where $\Phi$ is the Euler totient function.

This can be used to rewrite the shifted field at the origin as

$$\rho(\mathbf{r} = 0; \{\chi_i\}) =$$
$$2\sum_{m=1}^{n}[\cos\chi_m + \cos\chi_{m+2n} + \cos(\chi_m + \chi_{m+2n})]$$
$$+ 2\sum_{m=1}^{n}[\cos(\chi_m + \chi_{m+1})$$
$$+ \cos(\chi_{m+2n} + \chi_{m+2n+1})$$
$$+ \cos(\chi_m + \chi_{m+1} + \chi_{m+2n} + \chi_{m+2n+1})]. \qquad (56)$$

Each of the triplets of cosines in the two sums of equation (56) can be written in the form $\cos x + \cos y + \cos(x + y)$. This function has a minimum of $-3/2$ when both $x \equiv y \equiv \pm 2\pi/3$. However, this bound is unattainable for all $2n$ cosine triplets: If we set $\chi_m \equiv \chi_{m+2n} \equiv \pm 2\pi/3$ for $m = 1 \dots n$, so as to obtain the $-3/2$ minimum for all the triplets of the first sum, then no matter what the sign choices are, at least one of the triplets in the second sum must have phases of zero and therefore does not achieve the $-3/2$ minimum. This implies that $\phi_{\min} > -6n$, and so $\gamma > 2$.

---
[15] We note that, for $6n > 46$, there are additional inequivalent arrangements of wavevectors to consider, corresponding to distinct Bravais lattices, which we ignore here. See Mermin *et al.* (1987) for additional information.

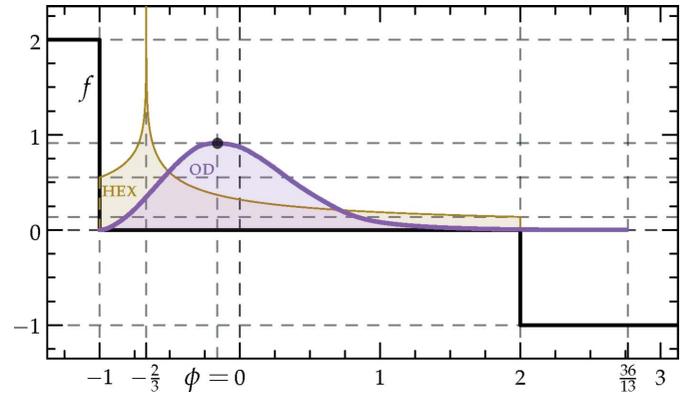

**Figure 22**
Density distribution of the two-scale octadecagonal quasicrystal: the black line is the local free-energy density from equation (57). Here $q = k_{18} \simeq 1.970$ and $\tilde\rho_1 = \tilde\rho_q = 1/13$. Note that the purple-colored octadecagonal density distribution extends into the positive values of $\phi > 2$, where the free-energy density is negative, making its overall free energy $\mathcal{F} \simeq -1.223 \times 10^{-3} < 0$. On the other hand, the density distribution of the single-scale hexagonal structure, which is plotted here for reference, cannot extend beyond $\phi = 2$ without running into the barrier at $\phi < -1$, which would force its free energy to become positive. This approach for forcing the quasicrystal structure to be the minimum free-energy state succeeds theoretically for all $6n$-fold quasicrystals, where $n \geq 2$, although they become increasingly fragile.

Indeed, numerical sampling for $\tilde\rho_1 = \tilde\rho_q$ shows that $\gamma = 3$, 3, $36/13 \simeq 2.769$ and $\sim 2.634$, for $n = 1\dots 4$, respectively. Asymptotically, $\gamma$ appears to decrease no faster than $2 + 2/n$.

## 8.2. A local free-energy density that stabilizes $6n$-fold quasicrystals

We set our local free-energy density to be

$$f(\phi) = \begin{cases} 2 & \phi < -1, \\ 0 & -1 \le \phi \le 2, \\ -1 & 2 < \phi, \end{cases} \qquad (57)$$

which is shown as the black lines in Figs. 1 and 22. Using the density distributions calculated in Section 6.4, it is not difficult to show that the uniform and single-scale lamellar and hexagonal phases must have non-negative free energies with this local free-energy function. In particular, because $\gamma_{\mathrm{HEX}} = 2$, the hexagonal phase cannot probe the negative $f(\phi > 2) = -1$ region without suffering greater free-energy penalties from the positive $f(\phi < -1) = 2$ region, as can be inferred graphically from Fig. 22. Furthermore, because the free-energy penalty is twice the free-energy decrease, the hexagonal phase is unable to reach negative free energies through negative values of $\tilde\rho$ either.

For the $6n$-fold two-scale structures considered above, we scale $\tilde\rho_1 = \tilde\rho_q$ until $\phi_{\min}$ reaches $-1$. Then, because $\gamma > 2$, $\phi_{\max}$ is also greater than two. This, together with the density distribution (36) and the local free-energy function (57), implies that the free energy $\mathcal{F}$ is negative for this structure. Thus, a $6n$-fold quasicrystal is the minimum free-energy state of the system. Indeed, the calculated free energies of the octadecagonal and icositetragonal quasicrystals are approximately $-1.223 \times 10^{-3}$ and $-6.093 \times 10^{-5}$, respectively. These free energies are simply the fraction of the density distribution above a density of two when the $\tilde{\rho}$'s are scaled such that $\phi_{min} = -1$, as can be seen graphically in Fig. 22.

Only some of the features of the contrived local free-energy density (57) are necessary for this stabilization to occur, and lower orders will be much more tolerant of imprecision than higher orders. For arbitrarily high orders, the flat region which includes $\phi = 0$ in the local free-energy functional is essential to this argument, as is some degree of favorable free energy for high values of $\phi$ and a somewhat greater free-energy penalty for sufficiently negative ones. These deviations from zero do not have to be sudden jumps. If the quasicrystalline $\gamma$ can only be proven to be greater than two, as we have done here for $6n$-fold structures, the ratio of the onset of these latter two effects must be exactly two, but if it can be shown to be even higher there will be some room for error.

Reproducing these results in the laboratory is likely to be challenging for at least three reasons:

(i) High length-scale selectivity will be required.

(ii) The thermodynamic stability of a stable state does not necessarily imply that it is kinetically accessible within a reasonable time frame.

(iii) Engineering an effective local free-energy density function like the one in equation (57) may be difficult. We suggest using a system similar to the interacting particles in the BDL model, where the vacuum constraint $\rho(\mathbf{r}) \geq -1$ implements the required barrier for negative concentrations relative to the average. A drop in the local free-energy density for sufficiently high concentrations could potentially be implemented through a kind of local phase change which sets in at a critical density, or some other highly nonlinear effect, such as the formation of oscillons (Umbanhowar et al., 1996; Arbell & Fineberg, 2000).

## 9. Closing remarks

In closing, we wish to emphasize the ease with which one can stabilize quasicrystals in rather simple isotropic models of interacting particles or their mean-field descriptions. It was appreciated from the outset that one needs to introduce multiple length scales into the interaction potentials of the constituent particles. Yet the ability to do so in a quantitatively predictive and controlled manner has only emerged in the last two decades, based on the understanding of how the multiple scales 'work together' to produce the targeted quasicrystalline structures.

The Faraday wave experiments of Edwards & Fauve (1993) led to the understanding of Lifshitz & Petrich (1997) that one needs to break the $\rho \to -\rho$ symmetry of the Landau free-energy expansion in order to allow the two length scales to couple via triad resonances or effective three-body interactions. This understanding was then generalized by Barkan et al. (2011) with their symmetry-breaking logarithmic entropy term.

Here, enabled by the density distribution method for calculating such non-polynomial free energies, we have come to an even deeper understanding that the breaking of $\rho \to -\rho$ symmetry favors the formation of structures with skewness. Indeed, quasicrystalline structures attain stability through the large skewness of their density distributions. Importantly, their extremes can be more lopsided than those of the hexagonal phase, which also takes advantage of its skewness to compete with the lamellar and uniform states. We have taken this idea to the extreme in Section 8 to design a local free energy which allows arbitrarily high-order quasicrystals to be stabilized.

Quasicrystals are stabilized by local free energies which take advantage of the unique skewed shape of quasicrystalline density distributions. Three-body interactions are responsible for this in the LP model, but any symmetry-breaking term may do the job.

### Acknowledgements

S. Savitz thanks Gil Refael, the Institute of Quantum Information and Matter, the Caltech Student–Faculty Programs office and Marcella Bonsall for their support. R. Lifshitz thanks Dean Petrich, Gilad Barak, Kobi Barkan, Yoni Mayzel, Michael Engel, Haim Diamant and Mike Cross for their fruitful collaboration on these problems over the years. We again extend our gratitude to Jiang et al. (2015) for sharing their data with us. Thanks to Marcus Bintz for pointing out the connections to the Kramers–Kronig relation and Morse theory in Section 6. The calculations in Sections 6 and 7 utilized the open-source computational geometry software library CGAL (The CGAL Project; https://www.cgal.org/).

### Funding information

Funding for this research was provided by: Israel Science Foundation (grant No. 1667/16).